\documentclass[12pt,preprint]{aastex}

\usepackage{amsmath}
\usepackage{graphics,graphicx}
\usepackage{subfig}
\usepackage{multirow}
\usepackage{array}
\usepackage{rotating}
\usepackage[T1]{fontenc}
\usepackage{textcomp}
\usepackage[hyphens]{url}
\usepackage{hyperref}

\begin{document}
%

\title{A Theory of Exoplanet Transits with Light Scattering}

\author{Tyler D. Robinson\altaffilmark{1,2,3}}
\affil{Department of Astronomy and Astrophysics, University of California, Santa Cruz, CA 95064, USA}
\email{tydrobin@ucsc.edu}

\altaffiltext{1}{Sagan Fellow}
\altaffiltext{2}{University of California, Santa Cruz, Other Worlds Laboratory}
\altaffiltext{3}{NASA Astrobiology Institute's Virtual Planetary Laboratory}

%
\begin{abstract}
Exoplanet transit spectroscopy enables the characterization of distant worlds, and will yield 
key results for NASA's {\it James Webb Space Telescope}.  However, transit spectra models 
are often simplified, omitting potentially important processes like refraction and multiple scattering.  
While the former process has seen recent development, the effects of light multiple scattering on 
exoplanet transit spectra has received little attention.  Here, we develop a detailed theory of 
exoplanet transit spectroscopy that extends to the full refracting and multiple scattering case.  We 
explore the importance of scattering for planet-wide cloud layers, where the relevant parameters are 
the slant scattering optical depth, the scattering asymmetry parameter, and the angular size of the 
host star.  The latter determines the size of  the ``target'' for a photon that is back-mapped from an 
observer.  We provide results that straightforwardly indicate the potential  importance of multiple 
scattering for transit spectra.  When the orbital distance is smaller than 10--20 times the stellar 
radius, multiple scattering effects for aerosols with asymmetry parameters larger than 0.8--0.9 can 
become significant.  We provide examples of the impacts of cloud/haze multiple scattering on transit 
spectra of a hot Jupiter-like exoplanet.  For cases with a forward and conservatively scattering cloud/haze, 
differences due to multiple scattering effects can exceed 200 ppm, but  shrink to zero at wavelength 
ranges corresponding to strong gas absorption or when the  slant optical depth of the cloud exceeds 
several tens.  We conclude with a discussion of  types of aerosols for which multiple scattering in 
transit spectra may be important.
\end{abstract}
%


%
\section{Introduction}

Transit spectroscopy \citep{seager&sasselov2000,brown2001,hubbardetal2001} 
is currently the leading technique for studying exoplanet atmospheric composition. 
Following the discovery of the first exoplanet atmosphere \citep{charbonneauetal2002}, 
transit observations have enabled the characterization of a number of different 
exoplanets for atmospheric molecular species, clouds, and/or hazes 
\citep{pontetal2008,swainetal2008,singetal2009,beanetal2010,fraineetal2014,
knutsonetal2014b}.  { Recent observational results have shown that hot Jupiter 
transit spectra demonstrate a complex continuum from clearsky to heavily clouded 
conditions \citep{singetal2016,stevensonetal2016}, and that cloudiness remains a 
key factor into the regime of lower mass planets \citep{lineetal2013b,
kreidbergetal2014a,knutsonetal2014a}.}

The field of exoplanet transit spectroscopy will be revolutionized with the 
anticipated launch of NASA's {\it James Webb Space Telescope} ({\it JWST}) 
in 2018 \citep{gardneretal2006}.  Over the course of the design five year 
mission for {\it JWST}, the observatory is expected to provide in-depth 
observations of many tens of transiting exoplanets \citep{beichmanetal2014}. 
Some of these observations will probe planets in the poorly understood 
2--4 Earth mass regime \citep{demingetal2009,batalhaetal2015}.  Excitingly, 
{\it JWST} may even be capable of characterizing temperate Earth-sized 
planets \citep{kaltenegger&traub2009,cowanetal2015,barstowetal2016}, 
though the ability to conduct such studies will depend largely on how well 
systematic noise sources can be constrained \citep{greeneetal2016}.

As the quality of transit spectrum observations continues to improve, 
so should models of exoplanet transits.  Thus, certain processes 
initially thought to be of second-order importance should be revisited 
and possibly added to modeling tools.  For example, atmospheric 
refraction, which was initially shown to be unimportant for the case 
of certain hot Jupiters \citep{hubbardetal2001}, has recently been 
shown to be critically important for understanding some terrestrial 
exoplanet spectra \citep{munozetal2012,betremieux&kaltenegger2014,
misraetal2014,betremieux&kaltenegger2015} and, possibly, gas giant 
transit spectra \citep{dalbaetal2015,betremieux2016}.  Additionally, 
refraction can affect the shape of an exoplanet transit lightcurve 
\citep{hui&seager2002,sidis&sari2010}.

Beyond refraction, another process that has seen little study with 
regards to exoplanet transits is light multiple scattering.  
\citet{hubbardetal2001} used a plane-parallel Monte Carlo scattering 
model to determine the significance of a molecular Rayleigh scattering 
``glow'' around the limb of a transiting exoplanet.  For the case of 
HD~209458b, these authors found that the contribution of Rayleigh 
scattering glow to the transit lightcurve is negligible.  Similar 
conclusions were reached by \citet{brown2001}, who used analytic 
arguments to deduce that multiple isotropic scatterings could have 
only a small impact on transit depths.

Since the study of \citet{hubbardetal2001}, scattering opacity in exoplanet 
transits has largely been treated as being equivalent to absorption 
opacity \citep[e.g.,][]{irwinetal2008,lineetal2013a,waldmannetal2015}.  
This equivalence cannot always hold.  For example, a very strong 
forward and conservatively scattering aerosol would only weakly 
attenuate a beam following a straight-line path from a stellar disk, passing 
through an exoplanet atmosphere, and traveling towards a distant 
observer.  This issue was plainly demonstrated by 
\citet{dekok&stam2012}, who used a Monte Carlo scattering model 
to study the transmission of a pencil beam through a cloud or haze 
layer with variable optical thickness and whose scatterers had asymmetry 
parameters of either 0, 0.9, or 0.98 (their Figure~3).  

{ In the work that follows, we do not seek to outline specific definitions for 
``hazes'' or ``clouds.''  In general, though, we use the former to refer to 
aerosols that form aloft and grow due to coagulation during the sedimentation 
process, thereby developing a distribution that may look something like 
Titan's tholin haze, which has a number density profile described (roughly) by an 
exponential with a constant scale height in Titan's upper atmosphere 
\citep{tomaskoetal2008}.  For clouds, we envision an aerosol distribution 
formed by lifting a gas to its condensation point, thereby developing a distinct 
cloud base with an overlying cloud deck whose thickness is controlled by 
mixing processes \citep[see, e.g.,][]{ackerman&marley2001}.  Advances in 
exoplanet and brown dwarf cloud models have been reviewed recently by 
\citet{marleyetal2013}.  The most popular tools include the {\tt Eddysed} 
model \citep{ackerman&marley2001}, where an upward diffusion of condensible 
vapor balances a downward sedimentation of aerosol particles, 
which has been successfully used to study both cloud and haze processes 
in transiting exoplanet atmospheres \citep{morleyetal2013}.  Additionally, the 
sophisticated microphysical cloud treatments of \citet{hellingetal2001} continue 
to see development and application to transiting exoplanets \citep{leeetal2016}.}

Thus, the question of the relative importance of scattering, refraction, 
and absorption remains largely unexplored for a wide portion of 
exoplanet parameter space.  Such studies have been hindered by the 
lack of available simulation tools, and also by a lack of a single, 
coherent theory of exoplanet transit spectroscopy.  Significant advances 
can be made by developing such a theory, and by building models that 
implement these ideas.

In this paper, we construct a theory of exoplanet transit spectroscopy 
that spans key fundamental integral relations through to its efficient, 
vectorized implementation.  Simplifications of the theory cover the 
geometric (i.e., straight-line) limit \citep[that is currently used in 
computationally-demanding spectral retrieval models; e.g.,][]{lineetal2011,
benneke&seager2012,barstowetal2012,leeetal2014,waldmannetal2015,
morleyetal2016} and cases that add refraction.  In general, though, we 
emphasize the ``full physics'' scenario, where a three-dimensional Monte 
Carlo model is used to incorporate scattering effects.  


Below, we introduce the concept of the ``path distribution'', which 
effectively separates the paths of photons (or rays) through a transiting 
exoplanet atmosphere from gaseous and/or aerosol absorption effects.  
As the former tends to vary weakly in wavelength, while the latter can 
vary strongly with wavelength, our approach is computationally efficient.
Following the presentation of these ideas, we validate our implementation 
of the theory against a number of trusted simulation tools.  We then 
give examples of the path distribution for a variety of different model 
atmospheres.  Finally, we use our tools to present the first simulated transit 
spectra of hot Jupiters with multiply-scattering clouds.

\section{Theory and Model Description}
\label{sec:theory}

Transit spectra typically probe low density regions of exoplanetary atmospheres.  Here, 
molecular absorption lines are relatively narrow, as the effects of pressure broadening 
are not dominant over Doppler broadening (as is the case in the deep atmosphere).  Thus, 
the gas opacity can vary by orders of magnitude over narrow spectral ranges.  By contrast, 
gaseous refractive indexes as well as gas and aerosol scattering properties tend to vary 
smoothly (and, sometimes, weakly) in wavelength.  Since refraction and scattering, which 
are more computationally intensive processes to simulate, influence the path of a 
photon (or ray) through an atmosphere, significant advances can be made by outlining a 
technique that separates the processes which influence photon trajectories from those 
which influence absorption.

We introduce the concept of the ray (or photon) atmospheric path distribution, 
$\mathcal{P}_{b}\!\left(h\right)$, where $b$ is the impact parameter of the ray and $h$ is 
altitude in the planetary atmosphere. We define the path distribution such that 
$\mathcal{P}_{b}\!\left(h\right)dh$ is the linear distance traversed by the ray at altitudes 
between $h$ and $h+dh$.  While $\mathcal{P}_{b}$ is dimensionless, it can be thought 
of as having units of km linear distance per km vertical distance.  The geometry 
of these parameters, for a simple case, is shown in Figure~\ref{fig:refract_geom}.  The 
Appendix contains additional discussion on the use of the path distribution.

In transit spectroscopy, the essential quantity is the attenuation of a ray along its 
path, as this distinguishes the opaque portions of the atmosphere (that block light 
from the stellar disk) from the transparent portions.  In 1-D atmospheric structure 
models, which are the variety most commonly used for exoplanets, the atmospheric 
opacity is only a function of altitude (or pressure) and wavelength.  Here, then, the 
optical depth along the path of a ray can be computed using the path distribution via, 
\begin{equation}
  \tau_{\lambda,b} = \int_{0}^{\infty} \alpha_{\lambda}(h) \mathcal{P}_{b}\!\left(h\right)dh \ ,
\label{eqn:tau_b}
\end{equation}
where $\alpha_{\lambda}$ is the atmospheric extinction (in units of inverse distance), 
and the integral can be taken to infinity as the extinction is zero at very large altitudes.
Since the layer vertical differential optical depth is defined as 
$d\tau_{\lambda}(h) =  \alpha_{\lambda}(h) dh$, this expression demonstrates that 
the path distribution can also be thought of as the linear optical depth encountered 
between $\tau_{\lambda}$ and $\tau_{\lambda} + d\tau_{\lambda}$.  Similar extensions 
exist for quantities like the column mass and number densities.  Note that 
Equation~\ref{eqn:tau_b} is critical, as the strongly wavelength-dependent extinction is 
separated from the ray path.  Using the standard definition of transmission,
\begin{equation}
  t_{\lambda}(b) = e^{ -\tau_{\lambda,b} } \ ,
\end{equation}
a simple transit spectrum in the pure absorbing limit (i.e., where all optical depth is 
taken as absorption optical depth) can then be computed by considering the light 
transmitted through concentric annuli on the planetary disk, each at their own impact 
parameter,
\begin{equation}
  \left( \frac{R_{\rm{p},\lambda}}{R_{\rm{s}}} \right)^{\! \! 2} =  \frac{2}{R_{\rm{s}}^2} \int_{0}^{\infty} \left[1 - t_{\lambda}(b) \right] b db \ ,
\label{eqn:transit_int}
\end{equation}
where $R_{\rm{p},\lambda}$ is the wavelength-dependent planetary radius, $R_{\rm{s}}$ 
is the stellar radius, and $ \left( R_{\rm{p},\lambda} / R_{\rm{s}} \right)^2$ is the transit 
depth.

In practice, model planetary atmospheres are defined on an altitude (or pressure) grid, 
and Equation~\ref{eqn:transit_int} is computed using a sum over a collection of impact 
parameters (and assuming that the planetary disk is opaque below some fiducial radius, 
$R_{\rm{p}}$, which is either at the surface or deep in the atmosphere).  In this case, 
the path distribution is a matrix, $\mathcal{P}_{i,j}$, where $\mathcal{P}_{i,j} \Delta h_j$ 
is the path traversed through atmospheric layer `$j$' (whose thickness is $\Delta h_j$) by 
a ray whose impact parameter is $b_{i}$.  At any given wavelength, the transmission is 
now a sum over $N_{\rm{lay}}$ atmospheric layers,
\begin{equation}
  t_{\lambda,i} = {\rm{EXP}}\! \left(  -\sum_{j=1}^{N_{\rm{lay}}} \alpha_{\lambda,j} \mathcal{P}_{i,j} \Delta h_j \right) 
                       = {\rm{EXP}}\! \left(  -\sum_{j=1}^{N_{\rm{lay}}} \Delta \tau_{\lambda,j} \mathcal{P}_{i,j} \right) \ ,
\label{eqn:trans_sum}
\end{equation}
where we have used the definition of the layer vertical differential optical depth, 
$\Delta \tau_{\lambda,j} = \alpha_{\lambda,j} \Delta h_j$.  A similar expression was used 
by \citet[][their Equations 5 and 7]{betremieux2016}, where these authors work in 
numerically-computed column number densities and molecular opacities, as compared 
to our dimensionless quantities.  Note that Equation~\ref{eqn:trans_sum} can be easily 
written in matrix notation as,
\begin{equation}
  \boldsymbol{t}_{\lambda} = \boldsymbol{1} - \boldsymbol{a}_{\lambda} 
                                          = {\rm{EXP}}\! \left( - \boldsymbol{\Delta \tau}_{\lambda} \cdot \boldsymbol{\mathcal{P}} \right) \ ,
\label{eqn:trans_matrix}
\end{equation}
where we have defined the absorptivity vector, $\boldsymbol{a}_{\lambda}$. The 
pure absorption transit spectrum derives from a sum over $N_{r}$ impact parameters,
\begin{equation}
  \left( \frac{R_{\rm{p},\lambda}}{R_{\rm{s}}} \right)^{\! \! 2} =  \frac{1}{R_{\rm{s}}^2} \left( R_{\rm{p}}^2  + 
                                                                                       2 \sum_{i=1}^{N_{r}} \left[1 - t_{\lambda,i} \right] b_{i} \Delta b_{i} \right) \ ,
\label{eqn:transit_sum}
\end{equation}
where $\Delta b_{i}$ is the thickness of the impact parameter gridpoint.  Continuing 
with the matrix notation, if we define a vector of annulus areas as 
$A_{i} = 2 \pi b_{i} \Delta b_{i}$, then the transit spectrum can be written 
simply as,
\begin{equation}
  \left( \frac{R_{\rm{p},\lambda}}{R_{\rm{s}}} \right)^{\! \! 2} =  \frac{1}{R_{\rm{s}}^2} \left( R_{\rm{p}}^2  + 
                                                                                       \frac{1}{\pi} \boldsymbol{a}_{\lambda} \cdot \boldsymbol{A} \right) \ .
\label{eqn:transit_vec}
\end{equation}

We briefly note that the transit spectra expressions given in 
Equations~\ref{eqn:transit_int}, \ref{eqn:transit_sum}, and \ref{eqn:transit_vec} are in 
the pure absorption limit, and assume that all rays trace back to the stellar disk, and 
that the star has uniform surface brightness.  While these are common assumptions 
when computing transit spectra for model atmospheres, we will now discuss the 
more general multiple scattering case.  The Appendix contains brief details about 
computing the path distribution in the geometric limit or in the case that refraction is 
considered.

\subsection{Multiple Scattering Path Distributions and Transmissions}
\label{section:scattering}

Cases that include clouds, especially strongly forward scattering clouds, 
require three-dimensional radiative transfer treatments to compute accurate transit 
spectra, and are well suited to Monte Carlo models.  Here, the path distributions for a 
number of photons, $N_{\rm{p}}$, are used to derive an average transmission 
for a grid of impact parameters.  The individual path distributions are determined 
using a Monte Carlo approach, and only consider the scattering optical depths 
($\tau_{\lambda,\rm{s}} = \tilde{\omega}_{0} \tau_{\lambda,\rm{e}}$, 
where $\tilde{\omega}_{0}$ is the single-scattering albedo, and a sub-script `s' 
indicates ``scattering'' while `e' indicates ``extinction'') since only 
$\tau_{\lambda,\rm{s}}$ affects the path of rays through the atmosphere.
Given the path distribution, $\boldsymbol{\mathcal{P}}_{m}$, for photon `$m$', 
the transmission for this photon is,
\begin{equation}
  \boldsymbol{t}_{\lambda,m} 
       = {\rm{EXP}}\! \left( - \boldsymbol{\Delta \tau}_{\lambda,a} \cdot \boldsymbol{\mathcal{P}}_{m} \right) \ ,
\label{eqn:phottrans}
\end{equation}
where $\boldsymbol{\Delta \tau}_{\lambda,a}$ is a vector of the 
wavelength-dependent layer differential absorption optical depths, and the 
transmission averaged over all photons is simply,
\begin{equation}
  \boldsymbol{\bar{t}}_{\lambda} 
       = \frac{\sum_{m=1}^{N_{\rm{p}}} \boldsymbol{t}_{\lambda,m} }{N_{\rm{p}}} \ .
\label{eqn:avgtrans}
\end{equation}

Execution of our Monte Carlo model follows a standard approach.  For a given 
impact parameter, $b_{i}$, a photon is directed into the atmosphere along the 
$-\hat{\boldsymbol{x}}$ direction in Figure~\ref{fig:refract_geom}, and then 
tracked through the atmosphere.  As this photon originated from the direction of 
the observer, we are performing a so-called ``backward'' Monte Carlo simulation.  
The optical distance the photon travels, either initially or following a scattering 
event, is determined by randomly sampling the scattering optical depth 
distribution, with, 
\begin{equation}
 \xi = \int_{0}^{\tau_{\rm{s}}} f(\tau) d\tau \ ,
\end{equation}
where $\xi$ is a random number between zero and one, and $f(\tau)d\tau$ 
is the probability that a photon scatters between $\tau$ and $\tau + d\tau$.  As 
$f(\tau)d\tau = e^{-\tau}d\tau$, we have, 
\begin{equation}
 \tau_{\rm{s}} = -\ln(1 - \xi) \ .
\end{equation}
For whichever layer (centered at $h_{j}$) the photon is currently in, the 
pathlength corresponding to $\tau_{\rm{s}}$ is,
\begin{equation} 
  s = \tau_{\rm{s}}/\alpha_{{\rm{s}},j} \ ,
\end{equation}
where $\alpha_{{\rm{s}},j}$ is the layer scattering extinction coefficient.  Given 
a pathlength, a photon location ($\boldsymbol{r} = x\hat{\boldsymbol{x}} + 
y\hat{\boldsymbol{y}} + z\hat{\boldsymbol{z}}$), and trajectory 
($\hat{\boldsymbol{\mu}} = \mu_{x}\hat{\boldsymbol{x}} + 
\mu_{y}\hat{\boldsymbol{y}} + \mu_{z}\hat{\boldsymbol{z}}$), the photon position 
is updated according to $x \rightarrow x + \mu_{x}s$, $y \rightarrow y + \mu_{y}s$, 
and $z \rightarrow z + \mu_{z}s$.

Depending on the size of $s$, the photon either experiences a scattering 
event within layer $j$, in which case $s/\Delta h_{j}$ is added to 
the path distribution for this photon (i.e., $\boldsymbol{\mathcal{P}}_{m}$), 
or the photon exits the layer before a scattering event occurs.  For the latter, 
$s$ must be compared to the distance to the layer boundaries.  Given the 
trajectory of the photon, and its current radial position, $r = \sqrt{x^2 + y^2 + z^2}$, 
the altitude of the photon along its path is,
\begin{equation}
  h(s) + R_{\rm{p}} = r(s) = \sqrt{ (x + \mu_{x}s)^2 + (y + \mu_{y}s)^2 + (z + \mu_{z}s)^2} \ .
\end{equation}
Thus, the quadratic equations (in $s$), 
\begin{equation}
  h(s_{j}) = h_{j} \pm \Delta h_{j}/2 \ ,
\end{equation}
give the distances to the layer boundaries.  When a photon travels 
to a layer boundary before a scattering event, $s_{j}/\Delta h_{j}$ 
is added to $\boldsymbol{\mathcal{P}}_{m}$, the cumulative 
optical depth experienced by the photon us updated as 
$\tau \rightarrow \tau + \alpha_{{\rm{s}},j}s_{j}$, and the photon is 
passed to the appropriate layer (either $j-1$ or $j+1$).  This process of 
randomly generating $\tau_{\rm{s}}$ and passing the photon through 
sequential layers is repeated until the photon either exits the atmosphere, 
or reaches the lower boundary of the model and is considered absorbed. 
(Note that, as with any transit spectrum model of gaseous worlds, the lower 
boundary must be set deep enough to not influence the simulated spectrum.)  
For photons that exit the atmosphere, the position and trajectory at exit 
determine whether or not the photon will intersect the stellar disk along a 
straight-line trajectory.

When a scattering event occurs, a scattering angle must be sampled 
from the scattering phase function and the photon trajectory must then 
be updated.  Assuming that the scattering phase function, $P$, is only 
a function of the cosine of a single angle, $\mu = \cos(\theta)$, then a 
randomly sampled scattering deflection angle is determined via,
\begin{equation}
  \xi = \int_{-1}^{\mu} P(\mu') d\mu' \ ,
\end{equation}
and an azimuthal scattering angle, $\phi$, is sampled uniformly, with,
\begin{equation}
  \phi = 2 \pi \xi \ .
\end{equation}
As the scattering angles are referenced from the propagation direction, 
a transformation must be done to convert the initial propagation 
direction, $\hat{\boldsymbol{\mu}}$, and the scattering angles into a 
new propagation direction, $\hat{\boldsymbol{\mu}}'$.  Standard 
expressions exist for completing this transformation 
\citep[e.g.,][their Equations 22 and 23]{witt1977}.

Refraction can be included in the Monte Carlo simulation.  As the 
photon moves along $s$, the path can be sub-divided, and a curvature 
applied to each smaller $\Delta s$.  At the photon height, the curvature,  
$r_{\rm{c}}$, is computed according to Equation~\ref{eqn:curvature}, and a 
deflection angle for the photon is determined from the refractive portion 
of Equation~\ref{eqn:dtheta_ds} (i.e., 
$\Delta \theta_{r} = \Delta s/r_{\rm{c}}$). The change in trajectory is 
referenced from $\hat{\boldsymbol{\mu}}$ and occurs in either the 
local upward or downward direction (depending on the sign of 
$r_{\rm{c}}$).  A new set of direction cosines is found by requiring, 
\begin{equation}
  \hat{\boldsymbol{\mu}} \cdot \hat{\boldsymbol{\mu}}' = \cos(\Delta \theta_{r}) \ ,
\end{equation}
\begin{equation}
  \left( \hat{\boldsymbol{\mu}} \times \hat{\boldsymbol{r}} \right) \cdot \hat{\boldsymbol{\mu}}' = 0 \ ,
\end{equation}
and
\begin{equation}
  \lvert \hat{\boldsymbol{\mu}}' \rvert = 1 \ ,
\end{equation}
where the first equation implies that the trajectory changes through 
$\Delta \theta_{r}$, the second equation forces the trajectory change 
to be either locally up or down, and the final equation ensures that 
the new direction of travel is a unit vector.

The computational efficiency of the Monte Carlo approach to multiple scattering 
in transit spectra outlined above can be increased in several ways.  First, for a 
given impact parameter, the Monte Carlo routine need only be executed if the 
straight-line scattering optical depth is a substantial fraction of the absorption 
optical depth.  (We choose a conservative cutoff at 
$\tau_{\lambda,\rm{s}} < 10^{-3} \tau_{\lambda,\rm{a}}$.) Also, as suggested 
by \citet{dekok&stam2012}, the number of photons used in the Monte Carlo 
simulation at a given impact parameter should be influenced by the straight-line 
transmission---these authors recommend the use of $10^{5}$ photons in 
transparent conditions and up to $10^{8}$ photons in opaque conditions.  
However, \citet{dekok&stam2012} used a traditional Monte Carlo radiative transfer 
approach, where photons are lost to absorption while passing through the 
atmosphere, thus driving the need for very large numbers of photons in their 
simulations.  As we have separated absorption from scattering in our model, 
(nearly) all of the photons we simulate in our Monte Carlo exit the atmosphere, so 
that we only require $10^{4}$--$10^{6}$ photons.  Even more dramatic efficiency 
gains can be made by noting that, since our Monte Carlo model only considers 
scattering optical depths, these simulations need only be recomputed if either the 
layer differential scattering optical depths or asymmetry parameters change 
significantly.  Thus, the individual path distribution matrices for each of the photons 
in our simulations can be saved and reused over a wide range of wavelengths in 
a high-resolution spectral grid.

Of course, running $N_{\rm{p}}$ transmission calculations (i.e., using 
Equation~\ref{eqn:phottrans}) at every wavelength in a high-resolution grid becomes 
computationally unfeasible if the grid is large.  Here, runtime reductions of factors of 
10--100 can be straightforwardly made by computing the analytic Jacobian of 
Equation~\ref{eqn:avgtrans}, 
${\partial \boldsymbol{\bar{t}}_{\lambda}}/{\partial \boldsymbol{\Delta \tau}_{\lambda,a}}$, 
with,
\begin{equation}
  \left( \frac{\partial \boldsymbol{\bar{t}}_{\lambda}}{\partial \boldsymbol{\Delta \tau}_{\lambda,a}} \right)_{\!\! i,j}
       =  \frac{1}{N_{\rm{p}}} \sum_{m=1}^{N_{\rm{p}}} \frac{\partial t_{\lambda,m}(b_{i})}{\partial \Delta \tau_{\lambda,a,j}} 
       = - \frac{1}{N_{\rm{p}}} \sum_{m=1}^{N_{\rm{p}}} \mathcal{P}_{m,i,j} \cdot t_{\lambda,m}(b_{i}) \ .
\label{eqn:jacobian}
\end{equation}
This Jacobian can be used to rapidly adapt $\boldsymbol{\bar{t}}_{\lambda}$ to 
changes in $\boldsymbol{\Delta \tau}_{\lambda,a}$, and our tests indicate that this 
approach is accurate with up to 25--50\% variations in level-dependent absorption optical 
depths.  Since a set of path distributions, transmissions, and Jacobians apply to 
wavelengths where the atmospheric optical property profiles are similar, the Monte 
Carlo approach outlined above is best implemented within a spectral mapping 
model \citep{westetal1990,meadows&crisp1996}.

We note that the many efficiency gains outlined above stem from our separate 
treatments of scattering (which influences the path distributions) and absorption 
(which influences the path-derived transmissions and Jacobians).  So far as we 
know, this treatment is a novel approach to Monte Carlo radiative transfer.  
Additionally, our approach to multiple scattering can be applied beyond exoplanet 
transits, and, for example, could be used to study scattering effects in occultation 
observations \citep[i.e., an extension of][]{dekok&stam2012}.

\subsection{Generalized Transit Spectra}

The transit spectra expressions given in Equations~\ref{eqn:transit_int}, 
\ref{eqn:transit_sum}, and \ref{eqn:transit_vec} are primarily useful 
in the geometric limit, since rays in refracting or scattering models are 
not guaranteed to trace back to the stellar disk.  To efficiently model 
spectra in these latter cases, we generalize the concept of the vector 
of annulus areas, $\boldsymbol{A}$.  Key variables used in the discussion 
below are visualized in Figure~\ref{fig:integration_geom}.

Ultimately, a transit spectrum is determined by comparing intensities 
integrated over the range of solid angles influenced by the planet  
($\Omega_{\rm{p}}$) in the case when the planet is present versus 
when only the star is considered (with corresponding intensities 
$I_{\rm{p},\lambda}$ and $I_{\rm{s},\lambda}$, respectively).  Thus, 
\begin{equation}
  \left( \frac{R_{\rm{p},\lambda}}{R_{\rm{s}}} \right)^{\! \! 2} = 
  \frac{\int_{\Omega_{\rm{p}}} \!\! I_{\rm{s,\lambda}} d\Omega - \int_{\Omega_{\rm{p}}} \!\! I_{\rm{p,\lambda}} d\Omega}
  {\int_{\Omega_{\rm{s}}} \!\! I_{\rm{s,\lambda}} d\Omega} \ ,
\end{equation}
where $\Omega_{\rm{s}}$ is the range of solid angles corresponding to 
the stellar disk.  Given the large distance $D$ between Earth and any 
exoplanetary system, so that $d\Omega=dA/D^2$, and taking the stellar 
intensity to be normalized such that,
\begin{equation}
  I_{0,\lambda} = \frac{\int_{\Omega_{\rm{s}}} \!\! I_{\rm{s,\lambda}} d\Omega}{ \ \ \Omega_{\rm{s}}} \ ,
\end{equation}
then,
\begin{equation}
  \left( \frac{R_{\rm{p},\lambda}}{R_{\rm{s}}} \right)^{\! \! 2} = 
  \frac{\int_{A_{\rm{p}}} \!\! {I_{\rm{s,\lambda}}}/{I_{0,\lambda}} dA - \int_{A_{\rm{p}}} \!\! {I_{\rm{p,\lambda}}}/{I_{0,\lambda}} dA}
  {\pi R_{\rm{s}}^{2} } \ .
\label{eqn:transit_general}
\end{equation}

The first integral in the numerator of Equation~\ref{eqn:transit_general} is 
simply a stellar limb darkening law integrated over the portion of the planetary 
disk that overlaps the stellar disk.  Using a polar coordinate system centered 
on the planetary disk, and letting $d$ be the distance from the center of the star 
to the center of the planet, this integral is then,
\begin{equation}
\int_{A_{\rm{p}}} \! \frac{I_{\rm{s,\lambda}}}{I_{0,\lambda}} dA = 
  \begin{cases}
    0 \ , &  d \geq R_{\rm{s}} + R_{\rm{p}} + h_{\rm{t}} \\
    \int_{r_{0}}^{R_{\rm{p}}+h_{\rm{t}}} \int_{-\theta_{0}(r)}^{\theta_{0}(r)} \frac{I_{\rm{s,\lambda}}}{I_{0,\lambda}} ( \mu_{\rm{s}} ) r d\theta dr \ , & d < R_{\rm{s}} + R_{\rm{p}} + h_{\rm{t}}  \ ,
  \end{cases}
\end{equation}
where $\mu_{\rm{s}} = \mu_{\rm{s}}(r,\theta)$ is the cosine of the angle of 
incidence on the stellar disk at coordinates $r$ and $\theta$, and
\begin{equation}
r_{0} = 
  \begin{cases}
    0 \ , &  d \leq R_{\rm{s}} \\
    d - R_{\rm{s}} \ , & d > R_{\rm{s}}  \ ,
  \end{cases}
\end{equation}
\begin{equation}
\theta_{0}(r) = 
  \begin{cases}
    \pi \ , &  d \leq R_{\rm{s}} - R_{\rm{p}} - h_{\rm{t}} \\
    \frac{d^2 + r^2 - R_{\rm{s}}^2}{2dr} \ , & d > R_{\rm{s}} - R_{\rm{p}} - h_{\rm{t}}  \ .
  \end{cases}
\end{equation}
However, as limb darkening is accounted for in standard transit 
observation data reduction procedures 
\citep[e.g.,][]{mandel&agol2002,kreidberg2015}, one may wish to omit 
the limb darkening law from the first integral in the numerator of 
Equation~\ref{eqn:transit_general}, leaving an analytic integral that can 
be performed by considering the geometry of two overlapping disks.

The second integral in the numerator of Equation~\ref{eqn:transit_general} 
contains the details associated with absorption, scattering, and refraction 
in the planetary atmosphere.  We define $\tilde{I}(r,\theta)$ as the 
background surface brightness mapped onto by an area element on the 
planet at $r$ and $\theta$.  This function is zero for rays that do not map 
back to the stellar disk, and, in the geometric limit, is equal to 
$I_{\rm{s,\lambda}}(\mu_{\rm{s}}(r,\theta))$ for portions of the planetary 
disk that overlap the star.  Given this definition, we then have,
\begin{equation}
  \int_{A_{\rm{p}}} \!  \frac{I_{\rm{p,\lambda}}}{I_{0,\lambda}} dA = 
  \frac{1}{I_{0,\lambda}} \int_{R_{\rm{p}}}^{R_{\rm{p}}+h_{\rm{t}}} \int_{-\pi}^{\pi} t_{\lambda}(r) \tilde{I}(r,\theta) r d\theta dr \ ,
\label{eqn:tdint}
\end{equation}
where we have assumed that $R_{\rm{p}}$ is set sufficiently deep so that 
$I_{\rm{p,\lambda}} = 0$ for $r < R_{\rm{p}}$.  Of course, it 
is straightforward to consider thermal emission from the planetary disk, 
which can be important at longer wavelengths \citep{kipping&tinetti2010}, 
by including the planet-to-star flux ratio for the planetary nightside on the 
right-hand side of Equation~\ref{eqn:transit_general}.

For efficient implementation, we define a grid on the atmospheric 
portion of the planetary disk in $N_{\theta}$ angular and $N_{r}$ radial points.  
An $N_{r} \times N_{\theta}$ area {\it matrix}, with 
$A_{i,j} = r_{i} \Delta r_{i} \Delta \theta_{j}$, need only be computed once, and 
the $N_{r} \times N_{\theta}$ background surface brightness mapping matrix, 
$\boldsymbol{\tilde{I}}$, is computed alongside the path distributions and 
transmissions.  Given these, we have
\begin{equation}
  \int_{A_{\rm{p}}} \!  \frac{I_{\rm{p,\lambda}}}{I_{0,\lambda}} dA = 
   \frac{1}{I_{0,\lambda}} \boldsymbol{t}_{\lambda} \cdot \left[ \left( \boldsymbol{\tilde{I}} \circ \boldsymbol{A} \right) \cdot \boldsymbol{1} \right] \ ,
\label{eqn:tdvec}
\end{equation}
where `$\circ$' indicates the Hadamard product, $\boldsymbol{1}$ is a 
vector of ones with length $N_{\theta}$, and, as before, 
$\boldsymbol{t}_{\lambda}$ is computed using either 
Equation~\ref{eqn:trans_matrix} or Equations~\ref{eqn:avgtrans} and 
\ref{eqn:jacobian}.  Critically, this approach still efficiently isolates the 
parameters that vary rapidly in wavelength, $\boldsymbol{t}_{\lambda}$, 
from the parameters influenced by geometry and ray paths---the background 
surface brightness mapping matrix need only be computed once in the 
geometric limit, and 1--2 times when refraction is included.  For the multiple 
scattering case, full Monte Carlo simulations need only be run at wavelength 
scales over which the atmospheric scattering properties vary, and 
transmissions need only be recomputed at wavelength scales over which 
their linear corrections via the analytic Jacobians become inaccurate.  
Integration can also be made more efficient by noting the symmetry about the 
line connecting the planet center to the star center.

The matrix $\boldsymbol{\tilde{I}}$ is straightforward to compute in the 
geometric limit, as rays trace straight lines back to either the stellar disk 
or not.  When refraction is included, as in the Appendix, 
a ray tracing is performed once for each gridpoint in $N_{r}$ ray impact 
parameters.  Then, given the exit location and direction for these rays, 
and the angular location, $\theta_{j}$, for each area element, simple 
geometry indicates which if the $N_{\theta}$ gridpoints have rays that 
map back to the stellar disk (and what the $\mu_{\rm{s}}$ value is for 
each of these rays).  In the multiple scattering case, $N_{\rm{p}}$ Monte Carlo 
simulations are run, and $\boldsymbol{\tilde{I}}$ is built up by averaging 
over the photons in each instance.  For each photon that exits the 
atmosphere, its trajectory leaving the top of the atmosphere is investigated 
to see if the stellar disk is intersected.  The geometry is then rotated 
through $\Delta \theta_{j}$, and the intersection is reinvestigated.  Thus, 
only $N_{\rm{p}}$ Monte Carlo simulations are needed to build up the 
average $N_{r} \times N_{\theta}$ background mapping matrix.

Figure~\ref{fig:simulation} shows the surface brightness in a full Monte Carlo 
simulation prior to integration over solid angle.  A haze with vertical 
optical depth of $\tau_{s}=0.01$ is placed above the 1~$\mu$bar pressure 
level following a linear power-law in pressure.  The atmosphere is isothermal 
at 1,500~K, the stellar and planetary size are appropriate for HD~189733b, 
and refraction and Rayleigh scattering are included (at 0.8~$\mu$m 
wavelength).  A secondary figure enhances the surface brightness off the 
stellar disk by a factor of 10.  { For the portion of the planet that is not 
overlapping the stellar disk, the sub-figure with enhanced surface 
brightness shows a thin ring due to forward scattered light.  Additional 
surface brightness structure in this ring is related to the scattering phase 
function---the region of the planetary limb that is at the greatest angular 
separation from the stellar disk less strongly samples the haze forward 
scattering peak.}

\subsection{Model Summary}

{ The ``path distribution'' formalism represents a framework that allows for the 
computation of transit spectra across a broad range of conditions.  Most existing 
transit spectrum models assume that rays travel along straight-line trajectories, 
and are extinguished from the line-of-sight path according to the extinction optical 
depth \citep[e.g.][]{lineetal2011,benneke&seager2012,barstowetal2012}.  
Equations~\ref{eqn:pathgeom}, \ref{eqn:trans_sum}, and \ref{eqn:transit_sum} apply 
in this simplified regime.

Slightly more sophisticated transit spectrum tools incorporate the physics of 
refraction through a ray tracing scheme \citep[e.g.][]{betremieux&kaltenegger2014,
misraetal2014}.  We outline a similar ray tracing approach to computing the path 
distribution for a refracting atmosphere in the Appendix.  For planets centered on 
their host stellar disk, only a single ray tracing need be performed.  This symmetry 
is broken as the planet moves away from the center of the disk, implying that 
different locations around the planetary disk may or may not refractively map back 
to the stellar disk.  Here, Equation~\ref{eqn:trans_sum} gives the (radially dependent) 
transmission, and Equation~\ref{eqn:tdint} or \ref{eqn:tdvec} describes the angular 
integration required to obtain a transit spectrum.

Extremely few models (or modeling approaches) exist that allow for the 
simulation of multiple scattering effects in transit spectra.  \citet{hubbardetal2001} 
describes a Monte Carlo model that treats the stratified limb of a transiting 
exoplanet as a set of non-interacting slabs.  Following these authors' investigation 
into Rayleigh scattering effects, this model has seen little use.  More recently, 
\citep{dekok&stam2012} constructed a realistic, three-dimensional Monte Carlo 
for investigating forward scattering effects in occultation and transit observations.  
This tool has not been used to predict exoplanet transit spectra that include the 
effects of forward scattering, though.  Our own three-dimensional Monte Carlo 
approach (detailed above), when paired with the path distribution formalism and 
with its use of analytic Jacobians, enables the calculation of high-resolution, 
multiple scattering transit spectra with computational efficiency gains of factors of 
10$^2$--10$^4$ over previous three dimensional models.}

To best enable the implementation of the theory outlined above, a stand-alone 
model, which we call {\tt scaTran}, has been made publicly available.  The software 
can be downloaded from \url{https://github.com/tdrobinson/scaTran}, or can 
be found by searching GitHub for ``scaTran.''

\section{Validation}
\label{sec:valid}

We validated our transit model, which we implement within the framework 
of the Spectral Mapping Atmospheric Radiative Transfer ({\tt SMART}) model 
\citep[developed by D.~Crisp;][]{meadows&crisp1996}, using a number of 
techniques and data sources.  First, with refraction and scattering optical 
depth omitted, we verified that our ray tracing routine (discussed in the 
Appendix) and the Monte Carlo approach return the path distribution in the 
geometric limit.  Then, with refraction included and scattering optical depth 
omitted, we showed that the ray tracing and Monte Carlo routines are in 
agreement.  Finally, to check our Monte Carlo approach with scattering 
included, we made the simple switch to a plane-parallel geometry (where 
height is only measured relative to the z-axis, not radially) and validated the 
output radiances against results from the {\tt DISORT} radiative transfer 
model \citep{stamnesetal1988} for a wide range of conditions.  
Figure~\ref{fig:mc_valid} shows the results from two of these experiments, 
plotting the top-of-atmosphere intensity (scaled by the incident flux) as a 
function of observer zenith angle.  Clouds with a given scattering optical 
depth and asymmetry parameter were placed over a perfectly absorbing 
surface, and a Henyey-Greenstein phase function was assumed.

To further validate our model, we performed model inter-comparisons. 
First, using a standard Earth atmospheric model \citep{mcclatcheyetal1972}, 
we compared our transit spectrum for an Earth-Sun twin system to the 
refracting model of \citet{misraetal2014}.  Note that the \citet{misraetal2014} 
model has been extensively validated against solar occultation data for Earth 
\citep{gunsonetal1996,irionetal2002} and observations of transmission 
through Earth's atmosphere during a lunar eclipse \citep{palleetal2009}.
Figure~\ref{fig:ref_valid} shows the result of this inter-comparison, where 
agreement (in effective transit height, equal to 
$R_{\rm{p},\lambda}-R_{\rm{p}}$) is to within the atmospheric model 
grid spacing.  Agreement improves if we use the coarser integration 
path lengths adopted by Misra~et~al. (i.e., 5~km).

Second, and finally, we applied our transit spectrum tool to a 
standardized hot Jupiter-like atmospheric model widely used for 
inter-comparison purposes.  This model has a planetary radius 
(at the 10 bar pressure level) of 1.16~$R_{\rm{J}}$, a planetary 
mass of 1.14~$M_{\rm{J}}$, a stellar radius of 0.78~$R_{\odot}$, 
and atmospheric volume mixing ratios of H$_2$, He, and 
H$_2$O of 0.85, 0.15, and $4\times10^{-4}$, respectively.  The 126 
model layers are placed evenly in log-pressure between 10 bar and 
10$^-9$ bar.  The atmosphere is isothermal at 1500~K.  
Figure~\ref{fig:geom_valid} compares our model, in the 
geometric limit, to output from the {\tt CHIMERA} retrieval suite 
\citep{lineetal2013a} for this standard case.  Agreement is within the 
20~ppm scatter seen in comparisons between other 
\citep{irwinetal2008,waldmannetal2015} transit spectrum tools 
(M.~R.~Line, personal communication).  We note that the offset in 
the Rayleigh scattering slopes between our model and the 
{\tt CHIMERA} calculation is due to differing approaches to 
computing Rayleigh scattering opacities which results in 
optical depths that differ at the 5\% level or less.

\section{Results}

We use the formalism outlined above to, first, compute the 
path distribution for Earth-like and hot Jupiter-like atmospheres 
for a range of conditions.  Following these instructive examples, 
we explore how scattering influences the transit depth associated 
with a single aerosol layer over a wide range of parameter space.  
Finally, we demonstrate the effects of thin, scattering clouds on 
a typical hot Jupiter transit spectrum.

\subsection{Path Distributions}

The aforementioned standard Earth and hot Jupiter atmospheric 
models can be used to demonstrate the path distribution for a range 
of conditions and assumptions.  To show the effects of clouds on 
the path distribution for the hot Jupiter atmosphere, we use the 
``cloud'' and ``haze'' differential optical depth profiles shown in 
Figure~\ref{fig:demo_dtau}.  The former is based on the shape of 
the aerosol mixing ratio profiles for a condensational cloud from the 
\citet{ackerman&marley2001} model, while the latter follows a 
power-law in altitude with scale height equal to the pressure scale 
height.  Both models are taken to have integrated vertical optical 
depth unity, and to be conservatively and moderately forward 
scattering ($g=0.9$).  A Henyey-Greenstein phase function is 
assumed for cloudy and hazy simulations.

Figure~\ref{fig:demo_geom} demonstrates the path distribution in the 
geometric limit for the standard hot Jupiter-like model atmosphere.  
Altitudes are measured relative to the 10 bar radius 
(at $1.16R_{\rm{J}}$).  This graph is interpreted by selecting an impact 
parameter, and using the color shading to interpret the path distribution 
(which, as stated earlier, can be thought of as an enhancement of the 
vertical optical depth).  For example, taking an impact parameter 
``altitude'' ($b - R_{\rm{p}}$) of 3000~km, the path distribution is zero 
for all atmospheric layers below this height, since rays with this impact 
parameter never pass through these deeper atmospheric layers.  The 
path distribution is then largest for layers near an altitude of 3000~km, 
as rays pass through these layers either horizontally or nearly 
horizontally.  Finally, for atmospheric layers well above 3000~km, rays 
pass through on much less extreme slant paths, so the path distribution 
is smaller (typically 5--15, for this particular model atmosphere).

The path distribution considering refraction for our cloud-free Earth 
atmosphere is shown in Figure~\ref{fig:demo_refract}.  For comparison, 
a case in the geometric limit is also presented.  For elements of the path 
distribution where the impact parameter altitude is 
near the surface, the path distribution deviates substantially from the 
geometric case due to the refractive bending of ray paths.  As an example, 
take an impact parameter altitude of 5~km.  In the geometric case, rays with 
this impact parameter would never reach altitudes below 5~km.  But, due to 
refraction, these rays are bent ``downward'' to probe deeper atmospheric 
layers.  Thus, the path distribution for the refracting case is now non-zero 
for atmospheric layers below 5~km.  Furthermore, the atmospheric layer 
that sees the most enhancement of optical depth (i.e., has the largest path 
distribution) is now slightly below 5~km, instead of being right at 5~km as in 
the geometric case.

Figures~\ref{fig:demo_ray} and \ref{fig:demo_clouds} show an average 
path distribution from our multiply-scattering Monte Carlo approach, with,
\begin{equation}
  \bar{\boldsymbol{\mathcal{P}}} 
   = \frac{1}{N_{\rm{p}}} \sum_{m=1}^{N_{\rm{p}}} \boldsymbol{\mathcal{P}}_{m} \ .
\end{equation}
We note that such an average path distribution is not used in our calculation 
of transit spectra, but is simply meant to indicate the characteristic path that 
scattered photons take through the atmosphere.  In the Figure~\ref{fig:demo_ray}
we only consider molecular Rayleigh scattering opacity (at 
0.55~$\mu$m wavelength) in our hot Jupiter-like model atmosphere. 
For Figure~ \ref{fig:demo_clouds}, we show the effects of 
our nominal haze and cloud models for the hot Jupiter-like case.  While 
both the haze and cloud have the same integrated optical depth, the 
haze path distribution has a less pronounced transition into the 
aerosol-affected atmospheric layers since the condensate cloud is 
concentrated into a more narrow range of altitudes/pressures.  Note the 
use of a logarithmic color contour scale to emphasize the impacts of 
scattering.

The path distributions for the scattering cases are less straightforward 
to interpret.  For the Rayleigh scattering case (Figure~\ref{fig:demo_ray}), 
and taking an impact parameter altitude of 800~km, we see that the path 
distribution is non-zero for altitudes below 800~km (but would be zero in 
the geometric case).  Here, a small fraction of photons have been 
scattered down to these deeper layers, and the small values of the path 
distribution at depth come at the (small) expense of the path distribution 
aloft.  Selecting a deeper impact parameter altitude of 200~km, we see 
that the path distribution is now greatest at large altitudes (where the 
path distribution would have been roughly 10 in the geometric case), and 
is small at altitudes below about 400--600~km.  Here, photons with a 
200~km impact parameter altitude are scattered at altitudes larger than 
about 400--600~km, so that relatively few of these photons can probe much 
deeper than this.

The haze case in Figure~\ref{fig:demo_clouds} can be interpreted 
similarly to the Rayleigh scattering case, although the range of 
impact parameters that are influenced by the haze is greater than that 
of the Rayleigh scattering case.  The condensate cloud case in 
Figure~\ref{fig:demo_clouds} appears distinct due to the well-defined 
cloud top.  Here, rays/photons with impact parameter altitudes larger 
than 1400~km never encounter the cloud, and the path distribution 
is the same as in the geometric case.  At altitudes just below this, 
though, the photons encounter the cloud, and are scattered before 
reaching atmospheric layers below the cloud (whose base is at about 
1200~km).  For these photons, their path distribution is concentrated at 
altitudes above the cloud and in the cloud itself.

\subsection{Exploration of the Importance of Scattering}

As most exoplanet transit spectrum models do not include multiple 
scattering, a key exercise is to explore the range of conditions for which 
scattering is expected to influence transit depth.  For this exploration, we 
note that the transit depth attributed to a narrow annulus on the planetary 
disk (i.e., at a single ray impact parameter) scales with 
$2 \pi b \Delta b / R_{\rm{s}}^{2}$.  Thus, in the absence of limb darkening, 
with the atmosphere transparent at radii larger than $b + \Delta b$, and with 
the planetary disk centered on the star, only three parameters will influence 
the annulus transit depth for a scattering scenario---the layer scattering 
optical depth, the aerosol scattering phase function, and the angular size of 
the stellar disk from the orbital location of the planet.

Figure~\ref{fig:tdepth_var} show the relative difference between a transit 
spectrum model in the geometric and pure absorption limits (where scattering 
optical depth is treated as absorption optical depth) and a multiple scattering 
model for a single annulus on the planetary disk.  Results are given for three 
different values of the host star angular size, which is determined by the ratio 
of the stellar radius to the planetary orbital distance ($a$).  These values span a 
host star angular size like that for Mercury and the Sun 
({ $a=0.39$~au}, so that $R_{\rm{s}}/a \sim 10^{-2}$) to that of a hot Jupiter-like 
planet orbiting at 0.05~AU from a Sun-like star ($R_{\rm{s}}/a \sim 10^{-1}$).  
Also, this range of host star angular sizes spans those for the Habitable Zones 
of M and K dwarf stars \citep{kopparapuetal2013}.

Color contours in Figure~\ref{fig:tdepth_var} are given as a function of 
layer scattering slant optical depth and the scattering asymmetry parameter.  No 
other opacity source is added to the layer.  For simplicity, a Henyey-Greenstein 
phase function is assumed \citep{henyey&greenstein1941}, as using a 
multi-parameter phase function would introduce additional variables to our phase 
space exploration.  Regardless, these results will still serve to indicate under 
which conditions scattering can become important.

\subsection{Thin Clouds and Hazes in Hot Jupiter Transits}

The standard hot Jupiter-like atmospheric model outlined in 
Section~\ref{sec:valid} provides a useful test case for exploring some important 
aspects of scattering in exoplanet transit spectra, especially when considering 
that planets near to their host stars are those for which scattering can have the 
largest effects.  The parameter space for such an exploration is extremely 
large, since the wavelength-dependent asymmetry parameter and scattering 
optical depth, as well as the cloud/haze vertical distribution, will all influence 
the transit spectrum.  Exploring all of this phase space is certainly beyond the 
scope of this manuscript, so we adopt a straightforward set of conditions.  
{ Specifically, we investigate the impact of general, isolated aerosol layers, placed 
at different pressure levels in the model atmosphere.  Rather than adopt a specific 
cloud or haze optical depth profile (e.g., Figure~\ref{fig:demo_dtau}), we simply distribute 
the aerosol optical depth uniformly over a single pressure scale height.}  Scattering 
optical depths and asymmetry parameters are assumed gray, the single-scattering 
albedo is taken to be unity (i.e., pure scattering clouds), and, as before, a 
Henyey-Greenstein scattering phase function is adopted.

Figure~\ref{fig:tdepth_thincloud_g0.95} shows transit spectra for strongly forward 
scattering cloud/haze particles ($g=0.95$) over the 1--2~$\mu$m wavelength range, 
which overlaps the accessible wavelength range for {\it Hubble}/WFC3 
\citep{kimbleetal2008}.  Figure~\ref{fig:tdepth_thincloud_g0.90} is similar, but for 
less strongly forward scattering cloud/haze particles ($g=0.90$).  Single-layer 
clouds were placed at pressures of 10$^{-4}$, 10$^{-3}$, 10$^{-2}$, and 
10$^{-1}$~bar, and spanned a vertical range of $d\ln p = 1$.  The clearsky 
case is shown in black, while pure absorbing and scattering cases are shown 
in gray and purple, respectively.  For each cloud pressure, three scenarios with 
different slant scattering optical depths are shown, with $\tau_{\rm{s}}$ equal 
to 1, 10, and 100.  This range of scattering optical depths spans the limits from 
a relatively optically thin aerosol layer to a relatively optically thick layer.

\section{Discussion}

The path distribution approach provides a coherent and computationally 
efficient method to computing transit spectra.  Furthermore, analyzing the 
path distribution can provide an understanding of which atmospheric layers 
can contribute information to rays emerging from the planetary disk at a 
given impact parameter.  For example, Figure~\ref{fig:demo_refract} 
demonstrates how refraction bends rays to probe deeper atmospheric layers than 
would be encountered in the geometric limit.  However, since the path distribution 
is intrinsic to just the planetary atmosphere, the other important effect of 
refraction---bending rays such that they do not strike the stellar disk---is not 
represented.  This is a strength of the path distribution approach, since only a 
single distribution would need to be computed for identical atmospheric 
models for planets orbiting different host stars.

Similarly, Figure~\ref{fig:demo_ray} shows how, on average, Rayleigh scattering will 
scatter a small fraction of photons at large impact parameters ``downward'' to probe 
deeper parts of the atmosphere.  Conversely, rays at small impact parameters no longer 
probe the deep atmosphere, as they are scattered at larger altitudes.  Of course, as 
Rayleigh scattering is not strongly forward scattering, it is unlikely that any significant 
fraction of the scattered photons will still intersect the stellar disk.

Figure~\ref{fig:demo_clouds} shows similar scattering effects to those 
in Figure~\ref{fig:demo_ray}.  However, the impact parameter where 
(roughly) deeper atmospheric layers become ineffectively probed depends 
on the vertical structure of the aerosols and, more specifically, where 
the slant scattering optical depth unity occurs.  We note, again, that these 
figures display average path distributions for our Monte Carlo simulations, 
and that transit spectra that include scattering must be computed using 
using the formalism developed in Section~\ref{section:scattering}.

Ultimately, when considering scattering, the brightness of any given 
annulus (at impact parameter $b$) on the planetary disk will depend on 
the transparency of atmospheric layers at radii larger than $b$, the slant 
scattering optical depth of the layer at $b$, the scattering phase function 
for the aerosols in this layer, and the angular size of the host star as seen 
from the planet.  For the situation where overlaying layers are transparent 
(implying that we can, thus, see down to a cloud), Figure~\ref{fig:tdepth_var} 
provides a rough guide to the circumstances where scattering can be 
important.  Here, one can use simple details from a forward model---
the angular size of the host star, the scattering optical thickness of a cloud 
layer, and cloud scattering asymmetry parameter---to determine if 
scattering can safely be ignored in a simulation.

In the case where the host star angular size is much smaller than would be 
at a distance of roughly 0.4~au from a Sun-like star 
(i.e., $R_{\rm{s}}/a \sim 10^{-2}$), Figure~\ref{fig:tdepth_var} shows that 
essentially any amount of scattering will prevent the path of a ray from tracing 
back to the stellar disk (which is a very small ``target'').  With 
$R_{\rm{s}}/a$ at roughly 1--5$\times10^{-2}$, which, for example, is 
appropriate for the Habitable Zones of late M dwarf stars, strongly forward 
scattering aerosols ($g > 0.9$) can have significant impacts on the transit 
depth due to an annulus for clouds whose slant optical depths are 
not very optically thick (i.e., $\tau_{s}$ less than 10).  This effect becomes 
quite dramatic for hot Jupiter-like conditions, where the host star is a relatively 
large ``target.''  Here, even modestly forward scattering aerosols ($g \sim 0.8$) 
can cause substantial variations as compared to the commonly-used pure 
absorption assumption.  This dependence on the asymmetry parameter 
explains why both \citet{hubbardetal2001} and \citet{brown2001} found that 
multiple scattering was unimportant for their presented hot Jupiter transit 
spectra.  As these authors considered only either Rayleigh scattering or 
isotropic phase functions (which both have $g=0$), scattering could 
effectively be treated as absorption.

Our adoption of the Henyey-Greenstein phase function is driven by the 
computational simplicity of this parameterization, and also the need to capture 
the process of forward scattering while minimizing the number of added 
parameters to our simulations.  Nevertheless, is has been shown that the 
Henyey-Greenstein phase function, when compared to Mie calculations, 
can underestimate the power in the forward scattering peak 
\citep{toublanc1996,boucher1998}, especially at wavelengths comparable to 
the particle size.  Thus, depending on the specifics of a given aerosol in an 
exoplanetary atmosphere, the results shown in Figure~\ref{fig:tdepth_var} 
may underestimate the increased transmission due to forward scattered 
light at a given asymmetry parameter (as compared to the pure absorption 
limit).

Figures~\ref{fig:tdepth_thincloud_g0.95} and \ref{fig:tdepth_thincloud_g0.90} 
further explore the importance of scattering in hot Jupiter transit spectra.  
For high-altitude clouds/hazes, these results show that differences between a 
pure absorption model versus a scattering model can approach (or exceed)  
200 ppm, which is larger than the typical uncertainties achieved in current 
observations with {\it Hubble} \citep{kreidbergetal2014a} or that are expected 
for {\it JWST} \citep{greeneetal2016}.  Very optically thick clouds (i.e., with slant 
scattering optical depths approaching 100) approach the pure absorption limit, 
and the difference between a scattering model and a pure absorption model for 
a thin cloud (i.e., scattering optical depth less than about unity) is small since the 
overall impact of the cloud on the transit spectrum is weak.  Similarly, the impact 
of scattering for a deep cloud or haze layer is small, as few wavelength regions 
are sensitive to the presence of this aerosol layer.  

In general, the impact of aerosol multiple scattering will depend on cloud 
configurations and optical properties, and this manuscript does not undertake the 
complex microphysical calculations needed to attempt predictions of cloud/haze 
distributions and compositions in exoplanet atmospheres---our goal is to highlight 
the conditions where scattering becomes an important consideration.  Nevertheless, 
the results in Figures~\ref{fig:tdepth_thincloud_g0.95} and 
\ref{fig:tdepth_thincloud_g0.90} indicate that forward scattering exoplanet clouds 
analogous to Earth's cirrus clouds (which are at higher altitudes and have optical 
depths less than roughly unity) may warrant a multiple scattering approach.  
Additionally, any forward scattering cloud with a scale height comparable to (or 
greater than) the pressure scale height could require a multiple scattering model 
to properly simulate a transit spectrum, as transit depth variations due to multiple 
scattering effects in this clouds will be comparable to those of molecular features 
(whose scale are typically set by the pressure scale height).

Differences between the scattering and pure absorption models for the hot 
Jupiter cases will also depend on the asymmetry parameter and single-scattering 
albedo.  For very strongly forward scattering aerosols (i.e., asymmetry parameters 
larger than our adopted value of 0.95), the transit spectrum will approach the 
clearsky limit (if the aerosols are weakly- or non-absorbing).  In essence, such 
clouds would be invisible, having no effect on the transit depth.  On the other hand, 
asymmetry parameters much smaller than roughly 0.8--0.9 will push the scattering transit 
spectra towards the pure absorption limit, since these scattered photons are 
increasingly unlikely to connect to the observer.  Similarly, as the 
single-scattering albedo is decreased (away from our adopted value of unity), the 
scattering spectra will approach the pure absorption case.  The extent of the influence 
of the single-scattering albedo depends on the average number of scatterings the 
photons experience along their path.  For example, for our $\tau_{\rm{s}}=10$ case, 
the photons are scattered of order ten times, implying a single-scattering albedo 
of less than about 0.9 would reduce the transmission to roughly 50\%.

Perhaps the most important message from the scattering studies above is the 
influence of the scattering asymmetry parameter on transit spectra of close-in 
exoplanets.  Aerosols whose scattering properties vary from strongly forward 
scattering to weakly forward scattering could impart features in a transit spectrum, 
as the cloud (or haze layer) will be more transparent where the particles have 
a larger asymmetry parameter.  Such signatures would appear along with   
absorption bands caused by particle vibrational modes \citep{wakeford&sing2015}
and cloud base features \citep{vahidiniaetal2014}.  Additionally, as transit spectra 
of forward scattering clouds would be reproduced by relatively thinner clouds in 
the pure absorption limit, transit spectral retrievals using pure absorption models 
could return cloud thickness (or number densities) that are biased low, where 
retrieved cloud optical depths would, instead, represent an effective optical depth 
that incorporates information about the optical properties of the cloud particles and 
the stellar angular size.

Whether or not aerosol multiple scattering will be important for any given 
exoplanet transit spectrum will depend on a number of parameters, including the 
slant optical thickness of any haze/cloud structure and the scattering asymmetry 
parameter (which will both, in general, depend on wavelength).  The planet must 
also be located near to its host star, which limits the range of relevant aerosols 
to those which form in warm and hot atmospheric conditions.  For example, 
asymmetry parameters for water droplets in the visible wavelength range are 
0.8--0.9 \citep{kokhanovsky2004}, and span 0.74--0.94 for ice crystals 
\citep[where crystals with plate-like morphologies are the most strongly forward 
scattering;][]{mackeetal1998}.  Thus, conditions may be appropriate for 
multiple scattering to influence the transit spectra of potentially habitable worlds 
around the coolest stars \citep[e.g., TRAPPIST-1][]{gillonetal2016}.  Looking 
beyond Earth in the Solar System, \citet{dekok&stam2012} (their Figure~1) 
show that a variety of ices, dusts, droplets, and fractal hazes are both 
non-absorptive and forward scattering (with $g \gtrsim 0.9$) below 2--3 $\mu$m.  
The wavelength range for these Solar System cases is especially relevant to 
transiting exoplanets around later-type stars that may be investigated by 
{\it Hubble}, {\it JWST}, the Fast Infrared Exoplanet Spectroscopy Survey 
Explorer concept \citep[FINESSE;][]{derooetal2012}, and/or the Atmospheric 
Remote-Sensing Infrared Exoplanet Large-survey concept 
\citep[ARIEL;][]{tinettietal2016}.

The composition, size distributions, and optical properties of aerosols in 
warm and hot exoplanet atmospheres are largely unknown.  For the hottest 
exoplanets, metal and silicate condensates may be expected to form 
\citep{lunineetal1989,marleyetal1999,burrowsetal2001}.  \citet{budajetal2015} 
investigated the optical properties of a large variety of metallic and 
silicon-bearing aerosols, and showed that many of these species can be 
strongly forward scattering over certain wavelength ranges.  Especially relevant 
cases are aluminum oxide, forsterite, enstatite, and pyroxine.

It is interesting to note that our results show that multiple scattering can be 
an important consideration for exoplanet transits when the angular size of 
the host star is relatively large, which is the opposite regime from where 
refraction has been shown to be important \citep{betremieux&kaltenegger2014,
misraetal2014,betremieux&kaltenegger2015}.  For refraction, a host star with 
small angular size implies that relatively little refractive bending is required 
to deflect a ray off the stellar disk, thereby setting a floor in the transit spectrum 
which is not captured in the commonly-used geometric, pure absorption limit.  
By comparison, only a single scattering (with even a strong forward peak) is 
required to deflect a ray from the stellar disk, which is then in agreement with 
the geometric, pure absorption limit.  Of course, the opposite of these statements 
is true for a star with large angular size.  In future work we plan to further explore 
the importance of refraction in transit observations of gas giants.

Finally, since this manuscript primarily emphasizes the development of a 
light scattering theory for exoplanet transits and the impacts of vertically 
thin cloud layers, a number of future studies will be undertaken.  These will 
include scattering effects in vertically more-extended cloud structures, and 
could also investigate the role of the assumed (or computed) particle 
scattering phase function.  Finally, the formalism outlined above can be used 
to study light scattering effects in time-resolved transit lightcurves.

\section{Conclusions}

We have detailed a new theory of exoplanet transit spectroscopy that includes 
the effects of light multiple scattering.  By effectively separating the path that 
photons take through an exoplanet atmosphere from the absorption processes 
of that atmosphere, the technique discussed in this manuscript yields models 
that are both physically rigorous and computationally efficient.  This approach, 
which relies on the so-called ``path distribution'' defined herein, can be extended 
to other areas of study, including (most straightforwardly) stellar occultations by 
planetary atmospheres.

When applying our validated scattering model to isolated cloud layers, models 
show that multiple scattering is most important for cases where the exoplanet host 
star is large in angular size as seen from the world (i.e., when the orbital distance 
is less than 10--20 times the stellar radius).  In these cases, multiple scattering 
by aerosols with asymmetry parameters larger than 0.8--0.9 can have substantial 
effects on the transmission of the cloud layer.  For all cases, differences between 
a multiple scattering model and a geometric pure absorption diminish for clouds 
(or hazes) with slant scattering optical depths approaching 100.

In an exploratory case of a conservatively and forward ($g=$0.90--0.95) 
scattering cloud/haze layer in the atmosphere of a hot Jupiter, differences in the 
transit depth from a multiple scattering model and a model in the pure absorption 
limit can exceed 200 ppm.  These differences are most pronounced when the 
cloud/haze is well above the level of gas absorption optical depth unity, and 
when the slant scattering optical depth is of order several to several tens.  Future 
work will further explore the situations where scattering in more extended exoplanet 
cloud (or haze) layers is key to understanding transit observations.

\acknowledgements
TR gratefully acknowledges support from NASA through the Sagan Fellowship 
Program executed by the NASA Exoplanet Science Institute. The results reported 
herein benefitted from collaborations and/or information exchange within NASA's 
Nexus for Exoplanet System Science (NExSS) research coordination network 
sponsored by NASA's Science Mission Directorate.  Certain essential tools used 
in this work were developed by the NASA Astrobiology Institute's Virtual Planetary 
Laboratory, supported by NASA under Cooperative Agreement No. NNA13AA93A.
The author thanks M.~Line and J.~Lustig-Yaeger for sharing model outputs for 
inter-comparison purposes, and J.~Fortney and R.~de~Kok for providing helpful 
feedback.  This manuscript is dedicated to the memory of Prince, who passed 
away during the completion of this project---``thank you for a funky time.''

\appendix
\section{Appendix: Uses of the Path Distribution}
Given the ray atmospheric path distribution as defined in Section~\ref{sec:theory}, 
a number of key quantities can be derived.  First, the pathlength (in km, for example) 
traversed between any two altitudes, $h_{1}$ and $h_{2}$, by a ray incident with 
impact parameter $b$ is simply,
\begin{equation}
  s_{b}\!\left(h_{1},h_{2}\right) = \lvert \int_{h_{1}}^{h_{2}} \mathcal{P}_{b}\!\left(h\right)dh \ \rvert \ ,
\end{equation}
where the absolute value forces all distances to be measured non-negative.  Then, if 
$h=0$ at the bottom of a terrestrial planetary atmosphere or at some reference pressure 
(e.g., 10 bar) for a gaseous world, and $h_{\rm{t}}$ is the effective top of the atmosphere, 
the total linear distance traverse through the entire atmosphere by the ray is,
\begin{equation}
  s_{b}\!\left(0,h_{\rm{t}}\right) = \int_{0}^{h_{\rm{t}}} \mathcal{P}_{b}\!\left(h\right)dh \ ,
\end{equation}
where we note that this integral would diverge if taken to infinite altitude.

The pathlength integral can also be written in terms of two pressure coordinates, 
$p_{1}$ and $p_{2}$, assuming the ideal gas law and hydrostatic equilibrium, 
\begin{equation}
  s_{b}\!\left(p_{1},p_{2}\right) = \lvert \int_{p_{2}}^{p_{1}} \mathcal{P}_{b}\!\left(p\right) \frac{R_{\rm{g}}T\!\left(p\right)}{mg} d\ln p \ \rvert
                                                =  \lvert \int_{p_{2}}^{p_{1}} \mathcal{P}_{b}\!\left(p\right) H\!\left(p\right) d\ln p \ \rvert \ ,
\end{equation}
where $R_{\rm{g}}$ is the universal gas constant, $T(p)$ is the atmospheric 
temperature profile, $g$ is the acceleration due to gravity, $m$ is the 
atmospheric mean molecular weight, and $H(p)$ is the pressure scale height.  
Also, given the definition of the path distribution, and the differential definition 
of  (path) column mass from the atmospheric mass density profile [$\rho(h)$],
\begin{equation}
  d\mathcal{M}_{\rm{c}} = \rho(h) ds \ ,
\end{equation}
the air mass encountered by a ray with impact parameter $b$ between two 
altitudes, $h_{1}$ and $h_{2}$, is given by,
\begin{equation}
  \mathcal{M}_{{\rm{c}},b}\!\left(h_{1},h_{2}\right) 
   = \lvert \int_{h_{1}}^{h_{2}} \rho(h) \mathcal{P}_{b}\!\left(h\right) dh \ \rvert \ .
\end{equation}
Similarly, a column (path) number density is,
\begin{equation}
  \mathcal{N}_{{\rm{c}},b}\!\left(h_{1},h_{2}\right) 
   = \lvert \int_{h_{1}}^{h_{2}} n(h) \mathcal{P}_{b}\!\left(h\right)dh \ \rvert \ ,
\end{equation}
where $n(h)$ is the number density profile.

\section{Appendix: Geometric Limit}
%


In the geometric limit, rays pass straight through the planetary atmosphere.  
Here, the path distribution can be determined analytically using geometric 
arguments.  For an atmospheric layer centered at $h$ with width $\Delta h$, and 
for a ray incident on the atmosphere with impact parameter $b$, the path 
distribution is given by
\begin{equation}
\mathcal{P}_{b}\!\left(h\right) = 
  \begin{cases}
    0 \ , & b \geq r +\Delta h/2 \\
    \frac{2}{\Delta h} \sqrt{ \left( r + \Delta h/2 \right)^2 - b^2} \ , &  r - \Delta h/2 < b < r + \Delta h/2 \\
    \frac{2}{\Delta h} \left[\sqrt{ \left( r + \Delta h/2 \right)^2 - b^2} - \sqrt{ \left( r - \Delta h/2 \right)^2 - b^2} \right] \ , & b \leq r - \Delta h/2 \ .  \\ 
  \end{cases}
\label{eqn:pathgeom}
\end{equation}
where we have defined $r = R_{\rm{p}} + h$ for conciseness.  Note that this 
expression is only independent of $\Delta h$ in the limit that this value is small 
when compared to $r$.  Performing a first-order expansion in $\Delta h$ yields,
\begin{equation}
\mathcal{P}_{b}\!\left(h\right) \approx \frac{2r}{\sqrt{r^2 - b^2}} \ , 
\end{equation}
which is in agreement with the linearized geometry discussed in 
\citet{fortney2005} (their Figure~1).  The geometric path distribution need only 
be computed once for a model atmosphere.  Thus, when paired with 
Equation~\ref{eqn:transit_vec}, the geometric approach can be executed with 
great computational efficiency.

\section{Appendix: Refraction-Only Cases}

When considering refraction, rays no longer travel on straight-line paths 
through the atmosphere.  In this case, the path distribution must be 
computed numerically.  Fortunately, as the refractive indexes for gases 
likely to be major atmospheric constituents vary weakly in wavelength, 
the path distribution need only be computed at a small number of 
spectral points when generating a transit spectrum.

Refraction will deflect the trajectory of a ray upward or downward, 
depending on the sign of the local atmospheric refractive index profile.
Here, the local curvature experienced by the ray, $r_{\rm{c}}$, is,
\begin{equation}
  \frac{1}{r_{\rm{c}}} = \sin\left( \theta_{r} \right) \frac{d\ln n_{\rm{ref}}}{dh} \ ,
\label{eqn:curvature}
\end{equation}
where $\theta_{r}$ is the zenith angle for the ray trajectory, and 
$n_{\rm{ref}}$ is the atmospheric refractive index.  The zenith angle is 
determined via,
\begin{equation}
  \cos( \theta_{r} ) = \hat{\boldsymbol{r}} \cdot \hat{\boldsymbol{\mu}} \ .
\end{equation}
Using pathlength as the integration variable, and according 
to the geometry shown in Figure~\ref{fig:refract_geom}, we have,
\begin{equation}
  \frac{dh}{ds} = \cos\left( \theta_{r} \right) \ ,
\end{equation}
\begin{equation}
  \frac{d\theta_{r}}{ds} = - \left[ \frac{\sin\left( \theta_{r} \right)}{r} + \frac{1}{r_{\rm{c}}} \right] \ ,
\label{eqn:dtheta_ds}
\end{equation}
\begin{equation}
  \frac{d\phi}{ds} = \frac{\sin\left( \theta_{r} \right)}{r} \ ,
\end{equation}
and
\begin{equation}
  \frac{d\omega}{ds} =  \frac{1}{r_{\rm{c}}} \ ,
\end{equation}
where $\phi$ is the polar angle measured from the  
$+\hat{\boldsymbol{x}}$ direction, and $\omega$ is the so-called 
refraction integral (which measures the deflection from the initial 
straight-line trajectory).

For a ray with initial impact parameter $b_{i}$, the path is determined by 
numerically integrating the expressions above using an algorithm like that 
described in \citet{vanderwerf2008}. While the ray is passing through the 
atmospheric layer at $h_{j}$, the increments of $\Delta s$ used in the path 
integration are divided by $\Delta h_{j}$ and added to $\mathcal{P}_{i,j}$.  
The path integration proceeds through all atmospheric layers until either the 
ray exits the atmosphere or the planetary surface is struck.  When the ray 
exits the atmosphere, the radial coordinate and trajectory are stored for later 
use in determining which rays map back to the stellar disk.  The straight-line 
trajectory the ray follows after exiting the atmosphere is defined by the total 
refraction angle at exit, which is equal to the direction cosine of the ray along 
the x-axis in Figure~\ref{fig:refract_geom} (i.e., 
$\hat{\boldsymbol{\mu}} \cdot \hat{\boldsymbol{x}} = \mu_{x} = \omega$).



\begin{thebibliography}{}

\expandafter\ifx\csname natexlab\endcsname\relax\def\natexlab#1{#1}\fi

\bibitem[{{Ackerman} \& {Marley}(2001)}]{ackerman&marley2001}
{Ackerman}, A.~S., \& {Marley}, M.~S. 2001, \apj, 556, 872

\bibitem[{{Barstow} {et~al.}(2012){Barstow}, {Aigrain}, {Irwin}, {Bowles},
  {Fletcher}, \& {Lee}}]{barstowetal2012}
{Barstow}, J.~K., {Aigrain}, S., {Irwin}, P.~G.~J., {et~al.} 2012, ArXiv
  e-prints, arXiv:1212.5020

\bibitem[{{Barstow} {et~al.}(2016){Barstow}, {Aigrain}, {Irwin}, {Kendrew}, \&
  {Fletcher}}]{barstowetal2016}
{Barstow}, J.~K., {Aigrain}, S., {Irwin}, P.~G.~J., {Kendrew}, S., \&
  {Fletcher}, L.~N. 2016, \mnras, 458, 2657

\bibitem[{{Batalha} {et~al.}(2015){Batalha}, {Kalirai}, {Lunine}, {Clampin}, \&
  {Lindler}}]{batalhaetal2015}
{Batalha}, N., {Kalirai}, J., {Lunine}, J., {Clampin}, M., \& {Lindler}, D.
  2015, ArXiv e-prints, arXiv:1507.02655

\bibitem[{Bean {et~al.}(2010)Bean, Kempton, \& Homeier}]{beanetal2010}
Bean, J.~L., Kempton, E. M.-R., \& Homeier, D. 2010, Nature, 468, 669

\bibitem[{Beichman {et~al.}(2014)Beichman, Benneke, Knutson, Smith, Lagage,
  Dressing, Latham, Lunine, Birkmann, Ferruit, {et~al.}}]{beichmanetal2014}
Beichman, C., Benneke, B., Knutson, H., {et~al.} 2014, Publications of the
  Astronomical Society of the Pacific, 126, 1134

\bibitem[{Benneke \& Seager(2012)}]{benneke&seager2012}
Benneke, B., \& Seager, S. 2012, The Astrophysical Journal, 753, 100

\bibitem[{{B{\'e}tr{\'e}mieux}(2016)}]{betremieux2016}
{B{\'e}tr{\'e}mieux}, Y. 2016, \mnras, 456, 4051

\bibitem[{{B{\'e}tr{\'e}mieux} \&
  {Kaltenegger}(2014)}]{betremieux&kaltenegger2014}
{B{\'e}tr{\'e}mieux}, Y., \& {Kaltenegger}, L. 2014, \apj, 791, 7

\bibitem[{B{\'e}tr{\'e}mieux \& Kaltenegger(2015)}]{betremieux&kaltenegger2015}
B{\'e}tr{\'e}mieux, Y., \& Kaltenegger, L. 2015, Monthly Notices of the Royal
  Astronomical Society, 451, 1268

\bibitem[{{Boucher}(1998)}]{boucher1998}
{Boucher}, O. 1998, Journal of Atmospheric Sciences, 55, 128

\bibitem[{Brown(2001)}]{brown2001}
Brown, T.~M. 2001, The Astrophysical Journal, 553, 1006

\bibitem[{{Budaj} {et~al.}(2015){Budaj}, {Kocifaj}, {Salmeron}, \&
  {Hubeny}}]{budajetal2015}
{Budaj}, J., {Kocifaj}, M., {Salmeron}, R., \& {Hubeny}, I. 2015, \mnras, 454,
  2

\bibitem[{{Burrows} {et~al.}(2001){Burrows}, {Hubbard}, {Lunine}, \&
  {Liebert}}]{burrowsetal2001}
{Burrows}, A., {Hubbard}, W.~B., {Lunine}, J.~I., \& {Liebert}, J. 2001,
  Reviews of Modern Physics, 73, 719

\bibitem[{{Charbonneau} {et~al.}(2002){Charbonneau}, {Brown}, {Noyes}, \&
  {Gilliland}}]{charbonneauetal2002}
{Charbonneau}, D., {Brown}, T.~M., {Noyes}, R.~W., \& {Gilliland}, R.~L. 2002,
  \apj, 568, 377

\bibitem[{{Cowan} {et~al.}(2015){Cowan}, {Greene}, {Angerhausen}, {Batalha},
  {Clampin}, {Col{\'o}n}, {Crossfield}, {Fortney}, {Gaudi}, {Harrington},
  {Iro}, {Lillie}, {Linsky}, {Lopez-Morales}, {Mandell}, \&
  {Stevenson}}]{cowanetal2015}
{Cowan}, N.~B., {Greene}, T., {Angerhausen}, D., {et~al.} 2015, \pasp, 127, 311

\bibitem[{{Dalba} {et~al.}(2015){Dalba}, {Muirhead}, {Fortney}, {Hedman},
  {Nicholson}, \& {Veyette}}]{dalbaetal2015}
{Dalba}, P.~A., {Muirhead}, P.~S., {Fortney}, J.~J., {et~al.} 2015, \apj, 814,
  154

\bibitem[{De~Kok \& Stam(2012)}]{dekok&stam2012}
De~Kok, R., \& Stam, D. 2012, Icarus, 221, 517

\bibitem[{{Deming} {et~al.}(2009){Deming}, {Seager}, {Winn}, {Miller-Ricci},
  {Clampin}, {Lindler}, {Greene}, {Charbonneau}, {Laughlin}, {Ricker},
  {Latham}, \& {Ennico}}]{demingetal2009}
{Deming}, D., {Seager}, S., {Winn}, J., {et~al.} 2009, \pasp, 121, 952

\bibitem[{{Deroo} {et~al.}(2012){Deroo}, {Swain}, \& {Green}}]{derooetal2012}
{Deroo}, P., {Swain}, M.~R., \& {Green}, R.~O. 2012, in \procspie, Vol. 8442,
  Space Telescopes and Instrumentation 2012: Optical, Infrared, and Millimeter
  Wave, 844241

\bibitem[{Fortney(2005)}]{fortney2005}
Fortney, J.~J. 2005, Monthly Notices of the Royal Astronomical Society, 364,
  649

\bibitem[{{Fraine} {et~al.}(2014){Fraine}, {Deming}, {Benneke}, {Knutson},
  {Jord{\'a}n}, {Espinoza}, {Madhusudhan}, {Wilkins}, \&
  {Todorov}}]{fraineetal2014}
{Fraine}, J., {Deming}, D., {Benneke}, B., {et~al.} 2014, \nat, 513, 526

\bibitem[{{Gardner} {et~al.}(2006){Gardner}, {Mather}, {Clampin}, {Doyon},
  {Greenhouse}, {Hammel}, {Hutchings}, {Jakobsen}, {Lilly}, {Long}, {Lunine},
  {McCaughrean}, {Mountain}, {Nella}, {Rieke}, {Rieke}, {Rix}, {Smith},
  {Sonneborn}, {Stiavelli}, {Stockman}, {Windhorst}, \&
  {Wright}}]{gardneretal2006}
{Gardner}, J.~P., {Mather}, J.~C., {Clampin}, M., {et~al.} 2006, \ssr, 123, 485

\bibitem[{{Gillon} {et~al.}(2016){Gillon}, {Jehin}, {Lederer}, {Delrez}, {de
  Wit}, {Burdanov}, {Van Grootel}, {Burgasser}, {Triaud}, {Opitom}, {Demory},
  {Sahu}, {Bardalez Gagliuffi}, {Magain}, \& {Queloz}}]{gillonetal2016}
{Gillon}, M., {Jehin}, E., {Lederer}, S.~M., {et~al.} 2016, \nat, 533, 221

\bibitem[{Greene {et~al.}(2016)Greene, Line, Montero, Fortney, Lustig-Yaeger,
  \& Luther}]{greeneetal2016}
Greene, T.~P., Line, M.~R., Montero, C., {et~al.} 2016, The Astrophysical
  Journal, 817, 17

\bibitem[{Gunson {et~al.}(1996)Gunson, Abbas, Abrams, Allen, Brown, Brown,
  Chang, Goldman, Irion, Lowes, {et~al.}}]{gunsonetal1996}
Gunson, M.~R., Abbas, M., Abrams, M., {et~al.} 1996, Geophysical Research
  Letters, 23, 2333

\bibitem[{{Helling} {et~al.}(2001){Helling}, {Oevermann}, {L{\"u}ttke},
  {Klein}, \& {Sedlmayr}}]{hellingetal2001}
{Helling}, C., {Oevermann}, M., {L{\"u}ttke}, M.~J.~H., {Klein}, R., \&
  {Sedlmayr}, E. 2001, \aap, 376, 194

\bibitem[{{Henyey} \& {Greenstein}(1941)}]{henyey&greenstein1941}
{Henyey}, L.~G., \& {Greenstein}, J.~L. 1941, \apj, 93, 70

\bibitem[{Hubbard {et~al.}(2001)Hubbard, Fortney, Lunine, Burrows, Sudarsky, \&
  Pinto}]{hubbardetal2001}
Hubbard, W., Fortney, J., Lunine, J., {et~al.} 2001, The Astrophysical Journal,
  560, 413

\bibitem[{Hui \& Seager(2002)}]{hui&seager2002}
Hui, L., \& Seager, S. 2002, The Astrophysical Journal, 572, 540

\bibitem[{{Irion} {et~al.}(2002){Irion}, {Gunson}, {Toon}, {Chang}, {Eldering},
  {Mahieu}, {Manney}, {Michelsen}, {Moyer}, {Newchurch}, {Osterman},
  {Rinsland}, {Salawitch}, {Sen}, {Yung}, \& {Zander}}]{irionetal2002}
{Irion}, F.~W., {Gunson}, M.~R., {Toon}, G.~C., {et~al.} 2002, \ao, 41, 6968

\bibitem[{{Irwin} {et~al.}(2008){Irwin}, {Teanby}, {de Kok}, {Fletcher},
  {Howett}, {Tsang}, {Wilson}, {Calcutt}, {Nixon}, \&
  {Parrish}}]{irwinetal2008}
{Irwin}, P.~G.~J., {Teanby}, N.~A., {de Kok}, R., {et~al.} 2008, \jqsrt, 109,
  1136

\bibitem[{Kaltenegger \& Traub(2009)}]{kaltenegger&traub2009}
Kaltenegger, L., \& Traub, W. 2009, The Astrophysical Journal, 698, 519

\bibitem[{{Kimble} {et~al.}(2008){Kimble}, {MacKenty}, {O'Connell}, \&
  {Townsend}}]{kimbleetal2008}
{Kimble}, R.~A., {MacKenty}, J.~W., {O'Connell}, R.~W., \& {Townsend}, J.~A.
  2008, in \procspie, Vol. 7010, Space Telescopes and Instrumentation 2008:
  Optical, Infrared, and Millimeter, 70101E

\bibitem[{{Kipping} \& {Tinetti}(2010)}]{kipping&tinetti2010}
{Kipping}, D.~M., \& {Tinetti}, G. 2010, \mnras, 407, 2589

\bibitem[{{Knutson} {et~al.}(2014{\natexlab{a}}){Knutson}, {Benneke}, {Deming},
  \& {Homeier}}]{knutsonetal2014a}
{Knutson}, H.~A., {Benneke}, B., {Deming}, D., \& {Homeier}, D.
  2014{\natexlab{a}}, \nat, 505, 66

\bibitem[{{Knutson} {et~al.}(2014{\natexlab{b}}){Knutson}, {Dragomir},
  {Kreidberg}, {Kempton}, {McCullough}, {Fortney}, {Bean}, {Gillon}, {Homeier},
  \& {Howard}}]{knutsonetal2014b}
{Knutson}, H.~A., {Dragomir}, D., {Kreidberg}, L., {et~al.} 2014{\natexlab{b}},
  \apj, 794, 155

\bibitem[{{Kokhanovsky}(2004)}]{kokhanovsky2004}
{Kokhanovsky}, A. 2004, Earth Science Reviews, 64, 189

\bibitem[{{Kopparapu} {et~al.}(2013){Kopparapu}, {Ramirez}, {Kasting}, {Eymet},
  {Robinson}, {Mahadevan}, {Terrien}, {Domagal-Goldman}, {Meadows}, \&
  {Deshpande}}]{kopparapuetal2013}
{Kopparapu}, R.~K., {Ramirez}, R., {Kasting}, J.~F., {et~al.} 2013, \apj, 765,
  131

\bibitem[{{Kreidberg}(2015)}]{kreidberg2015}
{Kreidberg}, L. 2015, \pasp, 127, 1161

\bibitem[{Kreidberg {et~al.}(2014)Kreidberg, Bean, D{\'e}sert, Benneke, Deming,
  Stevenson, Seager, Berta-Thompson, Seifahrt, \& Homeier}]{kreidbergetal2014a}
Kreidberg, L., Bean, J.~L., D{\'e}sert, J.-M., {et~al.} 2014, Nature, 505, 69

\bibitem[{{Lee} {et~al.}(2016){Lee}, {Dobbs-Dixon}, {Helling}, {Bognar}, \&
  {Woitke}}]{leeetal2016}
{Lee}, G., {Dobbs-Dixon}, I., {Helling}, C., {Bognar}, K., \& {Woitke}, P.
  2016, \aap, 594, A48

\bibitem[{{Lee} {et~al.}(2014){Lee}, {Irwin}, {Fletcher}, {Heng}, \&
  {Barstow}}]{leeetal2014}
{Lee}, J.-M., {Irwin}, P.~G.~J., {Fletcher}, L.~N., {Heng}, K., \& {Barstow},
  J.~K. 2014, \apj, 789, 14

\bibitem[{{Line} {et~al.}(2013){Line}, {Knutson}, {Deming}, {Wilkins}, \&
  {Desert}}]{lineetal2013b}
{Line}, M.~R., {Knutson}, H., {Deming}, D., {Wilkins}, A., \& {Desert}, J.-M.
  2013, \apj, 778, 183

\bibitem[{{Line} {et~al.}(2012){Line}, {Zhang}, {Vasisht}, {Natraj}, {Chen}, \&
  {Yung}}]{lineetal2011}
{Line}, M.~R., {Zhang}, X., {Vasisht}, G., {et~al.} 2012, \apj, 749, 93

\bibitem[{Line {et~al.}(2013)Line, Wolf, Zhang, Knutson, Kammer, Ellison,
  Deroo, Crisp, \& Yung}]{lineetal2013a}
Line, M.~R., Wolf, A.~S., Zhang, X., {et~al.} 2013, The Astrophysical Journal,
  775, 137

\bibitem[{{Lunine} {et~al.}(1989){Lunine}, {Hubbard}, {Burrows}, {Wang}, \&
  {Garlow}}]{lunineetal1989}
{Lunine}, J.~I., {Hubbard}, W.~B., {Burrows}, A., {Wang}, Y.-P., \& {Garlow},
  K. 1989, \apj, 338, 314

\bibitem[{{Macke} {et~al.}(1998){Macke}, {Francis}, {McFarquhar}, \&
  {Kinne}}]{mackeetal1998}
{Macke}, A., {Francis}, P.~N., {McFarquhar}, G.~M., \& {Kinne}, S. 1998,
  Journal of Atmospheric Sciences, 55, 2874

\bibitem[{{Mandel} \& {Agol}(2002)}]{mandel&agol2002}
{Mandel}, K., \& {Agol}, E. 2002, \apjl, 580, L171

\bibitem[{{Marley} {et~al.}(2013){Marley}, {Ackerman}, {Cuzzi}, \&
  {Kitzmann}}]{marleyetal2013}
{Marley}, M.~S., {Ackerman}, A.~S., {Cuzzi}, J.~N., \& {Kitzmann}, D. 2013, in
  {Comparative Climatology of Terrestrial Planets}, ed. S.~J. {Mackwell}, A.~A.
  {Simon-Miller}, J.~W. {Harder}, \& M.~A. {Bullock} ({Tucson}: {University of
  Arizona Press}), 367--391

\bibitem[{{Marley} {et~al.}(1999){Marley}, {Gelino}, {Stephens}, {Lunine}, \&
  {Freedman}}]{marleyetal1999}
{Marley}, M.~S., {Gelino}, C., {Stephens}, D., {Lunine}, J.~I., \& {Freedman},
  R. 1999, \apj, 513, 879

\bibitem[{{McClatchey} {et~al.}(1972){McClatchey}, {Fenn}, {Selby}, {Volz}, \&
  {Garing}}]{mcclatcheyetal1972}
{McClatchey}, R.~A., {Fenn}, R.~W., {Selby}, J.~E.~A., {Volz}, F.~E., \&
  {Garing}, J.~S. 1972, {Optical Properties of the Atmosphere (Third Edition)},
  Tech. rep., Air Force Cambridge Research Labs

\bibitem[{{Meadows} \& {Crisp}(1996)}]{meadows&crisp1996}
{Meadows}, V.~S., \& {Crisp}, D. 1996, \jgr, 101, 4595

\bibitem[{{Misra} {et~al.}(2014){Misra}, {Meadows}, \& {Crisp}}]{misraetal2014}
{Misra}, A., {Meadows}, V., \& {Crisp}, D. 2014, \apj, 792, 61

\bibitem[{Morley {et~al.}(2013)Morley, Fortney, Kempton, Marley, Vissher, \&
  Zahnle}]{morleyetal2013}
Morley, C.~V., Fortney, J.~J., Kempton, E. M.-R., {et~al.} 2013, The
  Astrophysical Journal, 775, 33

\bibitem[{{Morley} {et~al.}(2016){Morley}, {Knutson}, {Line}, {Fortney},
  {Thorngren}, {Marley}, {Teal}, \& {Lupu}}]{morleyetal2016}
{Morley}, C.~V., {Knutson}, H., {Line}, M., {et~al.} 2016, ArXiv e-prints,
  arXiv:1610.07632

\bibitem[{Mu{\~n}oz {et~al.}(2012)Mu{\~n}oz, Osorio, Barrena,
  Monta{\~n}{\'e}s-Rodr{\'\i}guez, Mart{\'\i}n, \& Pall{\'e}}]{munozetal2012}
Mu{\~n}oz, A.~G., Osorio, M.~Z., Barrena, R., {et~al.} 2012, The Astrophysical
  Journal, 755, 103

\bibitem[{Pall{\'e} {et~al.}(2009)Pall{\'e}, Osorio, Barrena,
  Monta{\~n}{\'e}s-Rodr{\'\i}guez, \& Mart{\'\i}n}]{palleetal2009}
Pall{\'e}, E., Osorio, M. R.~Z., Barrena, R., Monta{\~n}{\'e}s-Rodr{\'\i}guez,
  P., \& Mart{\'\i}n, E.~L. 2009, Nature, 459, 814

\bibitem[{Pont {et~al.}(2008)Pont, Knutson, Gilliland, Moutou, \&
  Charbonneau}]{pontetal2008}
Pont, F., Knutson, H., Gilliland, R., Moutou, C., \& Charbonneau, D. 2008,
  Monthly Notices of the Royal Astronomical Society, 385, 109

\bibitem[{Seager \& Sasselov(2000)}]{seager&sasselov2000}
Seager, S., \& Sasselov, D. 2000, The Astrophysical Journal, 537, 916

\bibitem[{{Sidis} \& {Sari}(2010)}]{sidis&sari2010}
{Sidis}, O., \& {Sari}, R. 2010, The Astrophysical Journal, 720, 904

\bibitem[{Sing {et~al.}(2009)Sing, D{\'e}sert, Lecavelier Des~Etangs,
  Ballester, Vidal-Madjar, Parmentier, Hebrard, \& Henry}]{singetal2009}
Sing, D., D{\'e}sert, J.-M., Lecavelier Des~Etangs, A., {et~al.} 2009,
  Astronomy and Astrophysics, 505, 891

\bibitem[{{Sing} {et~al.}(2016){Sing}, {Fortney}, {Nikolov}, {Wakeford},
  {Kataria}, {Evans}, {Aigrain}, {Ballester}, {Burrows}, {Deming},
  {D{\'e}sert}, {Gibson}, {Henry}, {Huitson}, {Knutson}, {Lecavelier Des
  Etangs}, {Pont}, {Showman}, {Vidal-Madjar}, {Williamson}, \&
  {Wilson}}]{singetal2016}
{Sing}, D.~K., {Fortney}, J.~J., {Nikolov}, N., {et~al.} 2016, \nat, 529, 59

\bibitem[{Stamnes {et~al.}(1988)Stamnes, Tsay, Jayaweera, Wiscombe,
  {et~al.}}]{stamnesetal1988}
Stamnes, K., Tsay, S.-C., Jayaweera, K., Wiscombe, W., {et~al.} 1988, Applied
  optics, 27, 2502

\bibitem[{{Stevenson}(2016)}]{stevensonetal2016}
{Stevenson}, K.~B. 2016, \apjl, 817, L16

\bibitem[{{Swain} {et~al.}(2008){Swain}, {Vasisht}, \&
  {Tinetti}}]{swainetal2008}
{Swain}, M.~R., {Vasisht}, G., \& {Tinetti}, G. 2008, \nat, 452, 329

\bibitem[{{Tinetti} {et~al.}(2016){Tinetti}, {Drossart}, {Eccleston},
  {Hartogh}, {Heske}, {Leconte}, {Micela}, {Ollivier}, {Pilbratt}, {Puig},
  {Turrini}, {Vandenbussche}, {Wolkenberg}, {Pascale}, {Beaulieu}, {G{\"u}del},
  {Min}, {Rataj}, {Ray}, {Ribas}, {Barstow}, {Bowles}, {Coustenis}, {Coud{\'e}
  du Foresto}, {Decin}, {Encrenaz}, {Forget}, {Friswell}, {Griffin}, {Lagage},
  {Malaguti}, {Moneti}, {Morales}, {Pace}, {Rocchetto}, {Sarkar}, {Selsis},
  {Taylor}, {Tennyson}, {Venot}, {Waldmann}, {Wright}, {Zingales}, \&
  {Zapatero-Osorio}}]{tinettietal2016}
{Tinetti}, G., {Drossart}, P., {Eccleston}, P., {et~al.} 2016, in \procspie,
  Vol. 9904, Society of Photo-Optical Instrumentation Engineers (SPIE)
  Conference Series, 99041X

\bibitem[{Tomasko {et~al.}(2008)Tomasko, Doose, Engel, Dafoe, West, Lemmon,
  Karkoschka, \& See}]{tomaskoetal2008}
Tomasko, M., Doose, L., Engel, S., {et~al.} 2008, Planetary and Space Science,
  56, 669

\bibitem[{{Toublanc}(1996)}]{toublanc1996}
{Toublanc}, D. 1996, Applied optics, 35, 3270

\bibitem[{{Vahidinia} {et~al.}(2014){Vahidinia}, {Cuzzi}, {Marley}, \&
  {Fortney}}]{vahidiniaetal2014}
{Vahidinia}, S., {Cuzzi}, J.~N., {Marley}, M., \& {Fortney}, J. 2014, \apjl,
  789, L11

\bibitem[{van~der Werf(2008)}]{vanderwerf2008}
van~der Werf, S. 2008, Applied optics, 47, 153

\bibitem[{{Wakeford} \& {Sing}(2015)}]{wakeford&sing2015}
{Wakeford}, H.~R., \& {Sing}, D.~K. 2015, \aap, 573, A122

\bibitem[{{Waldmann} {et~al.}(2015){Waldmann}, {Tinetti}, {Rocchetto},
  {Barton}, {Yurchenko}, \& {Tennyson}}]{waldmannetal2015}
{Waldmann}, I.~P., {Tinetti}, G., {Rocchetto}, M., {et~al.} 2015, \apj, 802,
  107

\bibitem[{{West} {et~al.}(1990){West}, {Crisp}, \& {Chen}}]{westetal1990}
{West}, R., {Crisp}, D., \& {Chen}, L. 1990, \jqsrt, 43, 191

\bibitem[{{Witt}(1977)}]{witt1977}
{Witt}, A.~N. 1977, \apjs, 35, 1

\end{thebibliography}


\newpage

\section{Tables and Figures}
%
\begin{table}[ht]
  \centering
  \scriptsize
  {\bf Table 1. Symbol Usage} \\
  \vspace{2mm}
  \begin{tabular}{c l}
    \hline
    \hline
    Symbol &  Description \\
    \hline
   $\boldsymbol{A}$         &   vector/array of area elements \\
            $a$                      &   planet-star orbital distance \\ 
    $a_{\lambda}$             &   absorptivity along a ray path \\
 $\alpha_{\lambda}$        &   atmospheric extinction coefficient \\
          $b$                        &   impact parameter \\
          $d$                        &   projected separation between planet center and star center \\
          $g$                        &   scattering asymmetry parameter, acceleration due to gravity \\
$H = R_{\rm{g}}T/mg$     &   pressure scale height \\
   $h$, $h_{\rm{t}}$           &   altitude, altitude at effective top of atmosphere \\
 $I_{\rm{s},\lambda}$       &   stellar surface brightness \\
 $I_{\rm{p},\lambda}$       &   surface brightness on planetary disk \\
 $I_{\rm{0},\lambda}$       &   disk-averaged stellar surface brightness \\
        $\tilde{I}$                  &  background to planetary disk surface brightness mapping \\
          $m$                        &   atmospheric mean molecular weight \\
   $\mu = \cos(\theta)$      &   cosine of scattering angle \\
     $\mu_{\rm{s}}$            &   cosine of angle of incidence on stellar disk \\
$\hat{\boldsymbol{\mu}}$ &   direction of photon/ray travel \\
       $N_{r}$                     &   number of impact parameters in model grid \\
     $N_{\rm{p}}$               &   number of photons in a Monte Carlo simulation \\
   $N_{\rm{lay}}$              &   number of layers in atmospheric model \\
    $n_{\rm{ref}}$              &   atmospheric refractive index \\
       $P(\mu)$                  &   scattering phase function \\
 $\mathcal{P}_{b}$, $\boldsymbol{\mathcal{P}}$                                                                    &   atmospheric path distribution \\
          $p$                        &   pressure \\
        $\phi$                      &   polar angle when ray tracing, scattering azimuth angle \\
    $R_{\rm{E}}$               &   Earth radius \\
    $R_{\rm{g}}$               &   universal gas constant \\
    $R_{\rm{J}}$               &   Jupiter radius \\
    $R_{\rm{p}}$               &   fiducial planetary radius \\
$R_{\rm{p},\lambda}$     &   wavelength-dependent planetary radius \\
    $R_{\rm{s}}$               &   stellar radius \\
          $r$                        &   radial distance from planet center \\
$r_{0}$, $\theta_{0}$      &   integration bounds for stellar brightness/area \\
    $r_{\rm{c}}$                &   ray curvature due to refraction \\
       $s_{b}$                    &  ray path \\
      $t_{\lambda}$           &  transmission along ray path \\
$\boldsymbol{\bar{t}}_{\lambda}$                                                                                        & Monte Carlo average transmission vector \\
${\partial \boldsymbol{\bar{t}}_{\lambda}}/{\partial \boldsymbol{\Delta \tau}_{\lambda,a}}$ & Monte Carlo average transmission Jacobian matrix \\
     $\Delta\tau$              &  layer vertical optical depth \\
         $\tau$                    &  optical depth along ray path or slant optical depth \\
      $\theta_{r}$              &  local zenith angle for ray propagation direction \\
         $\theta$                 &  scattering deflection angle, polar angle in disk integration \\
$\Omega$, $d\Omega$ &  solid angle, differential solid angle \\
      $\omega$                 & refraction integral/angle \\
 $\tilde{\omega}_{0}$     &  single-scattering albedo \\
           $\xi$                    &  random number between zero and one \\
    \hline
  \end{tabular}
\end{table}
\clearpage
\begin{figure}
  \centering
  \includegraphics[trim = 0mm 0mm 0mm 0mm, clip, width=6.2in]{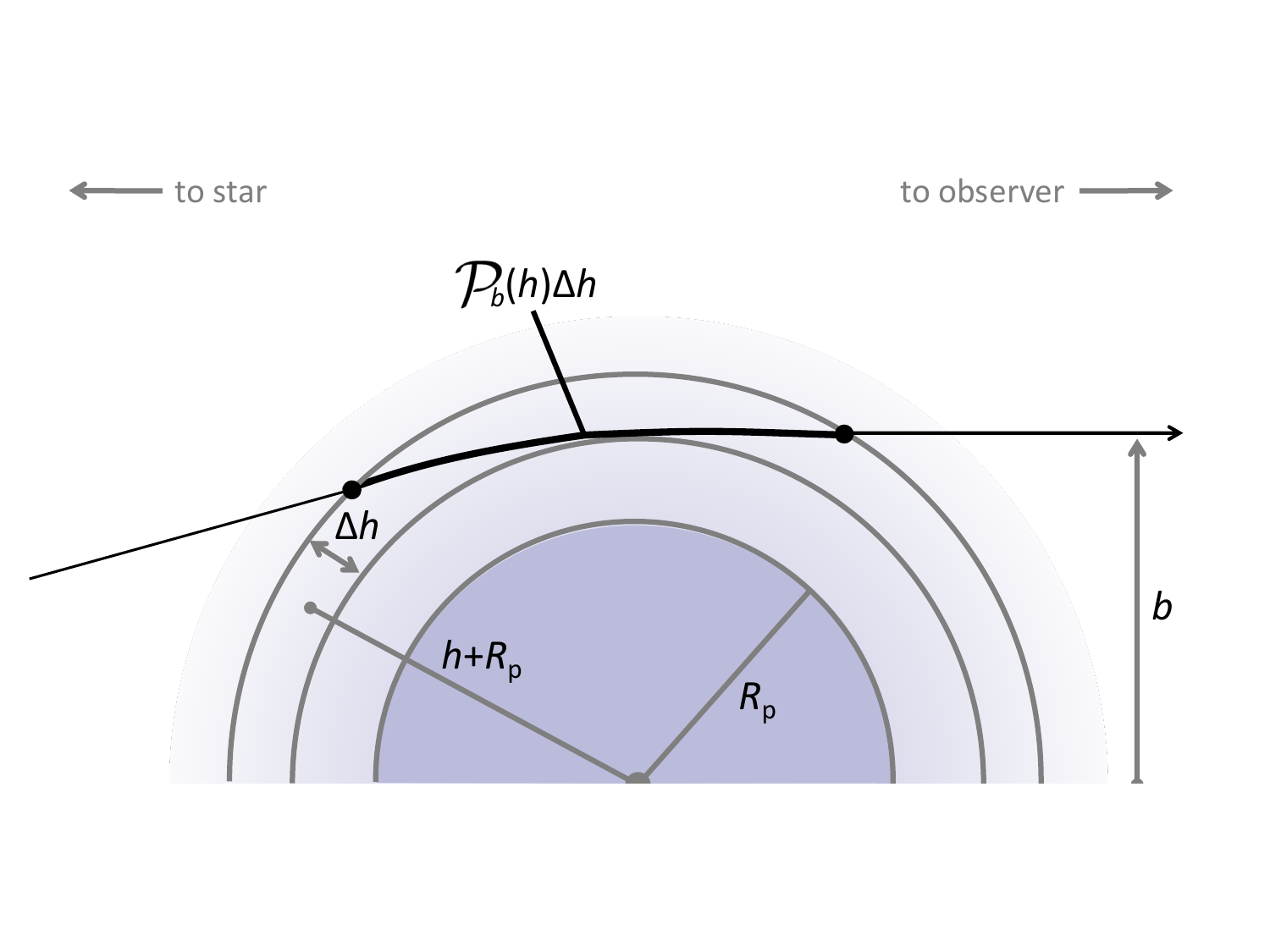}
  \caption{Visualization of the path distribution for a ray incident exiting a planetary atmosphere 
                with impact parameter $b$ and passing through an atmospheric layer of width 
               $\Delta h$ centered at altitude $h$.}
  \label{fig:geom_path}
\end{figure}
\clearpage
\begin{figure}
  \centering
  \includegraphics[trim = 0mm 0mm 0mm 0mm, clip, width=6.2in]{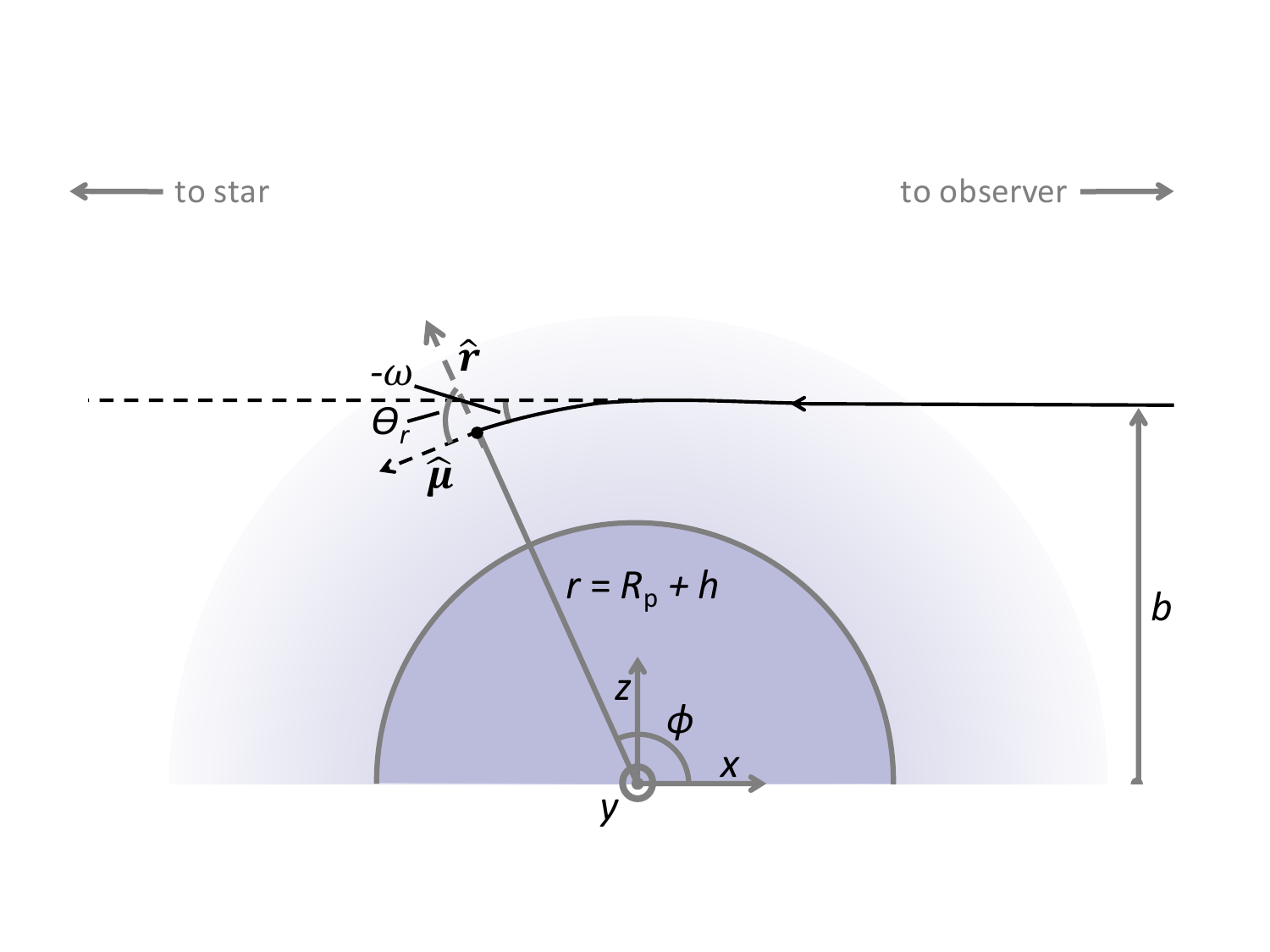}
  \caption{Visualization of key parameters in computing the path distribution in the
               non-geometric limit case.  Note how ray direction is reversed, as paths 
               are traced backwards to investigate whether they strike the stellar disk.}
  \label{fig:refract_geom}
\end{figure}
\clearpage
\begin{figure}
  \centering
  \includegraphics[trim = 0mm 0mm 0mm 0mm, clip, width=6.2in]{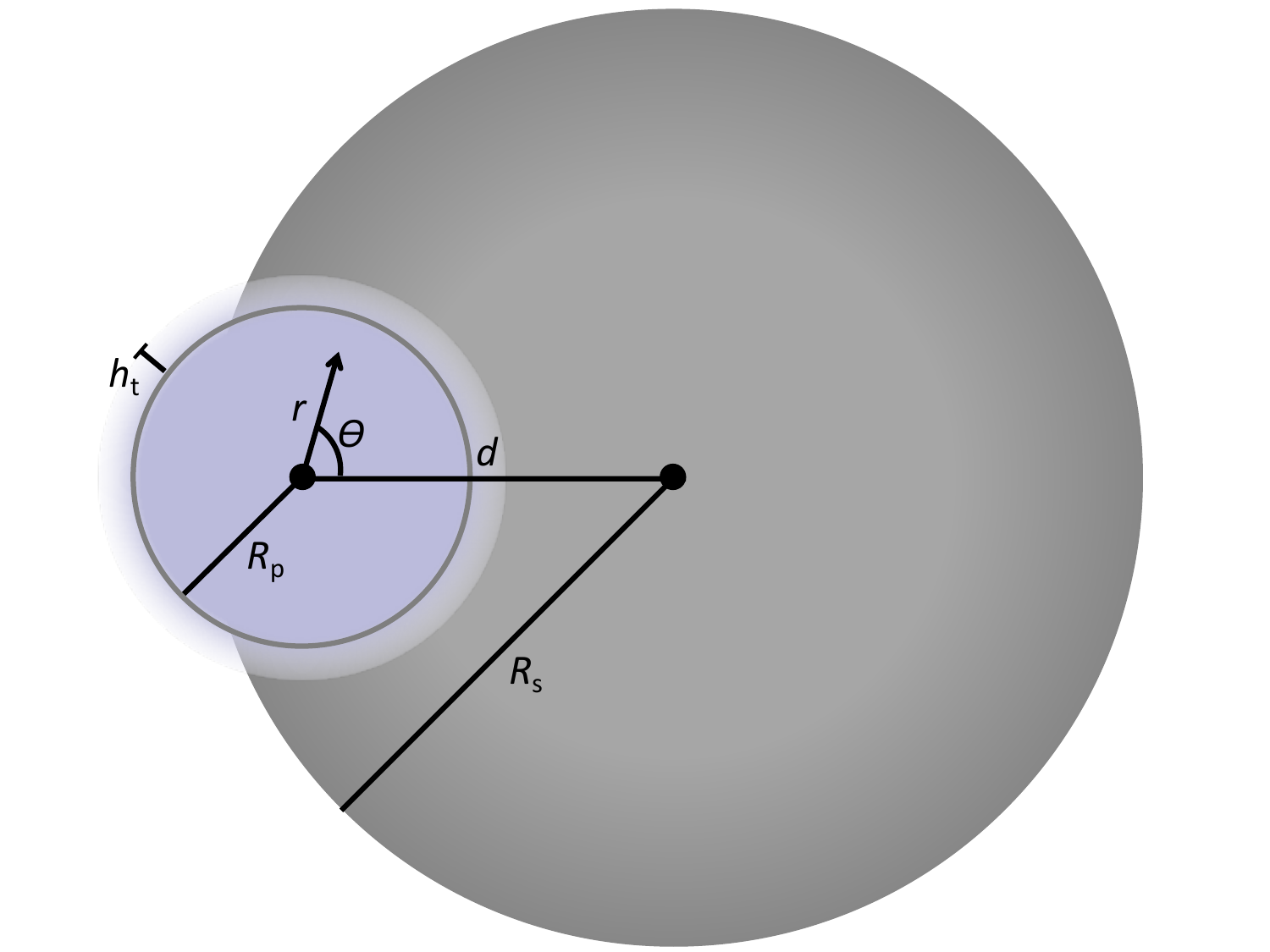}
  \caption{Visualization of geometry for integrating planetary and stellar surface brightness 
                when generating a transit spectrum.}
  \label{fig:integration_geom}
\end{figure}
\clearpage
\begin{figure}
  \centering
  \begin{tabular}{cc}
    \includegraphics[trim = 0mm 0mm 0mm 0mm, clip, width=3.1in]{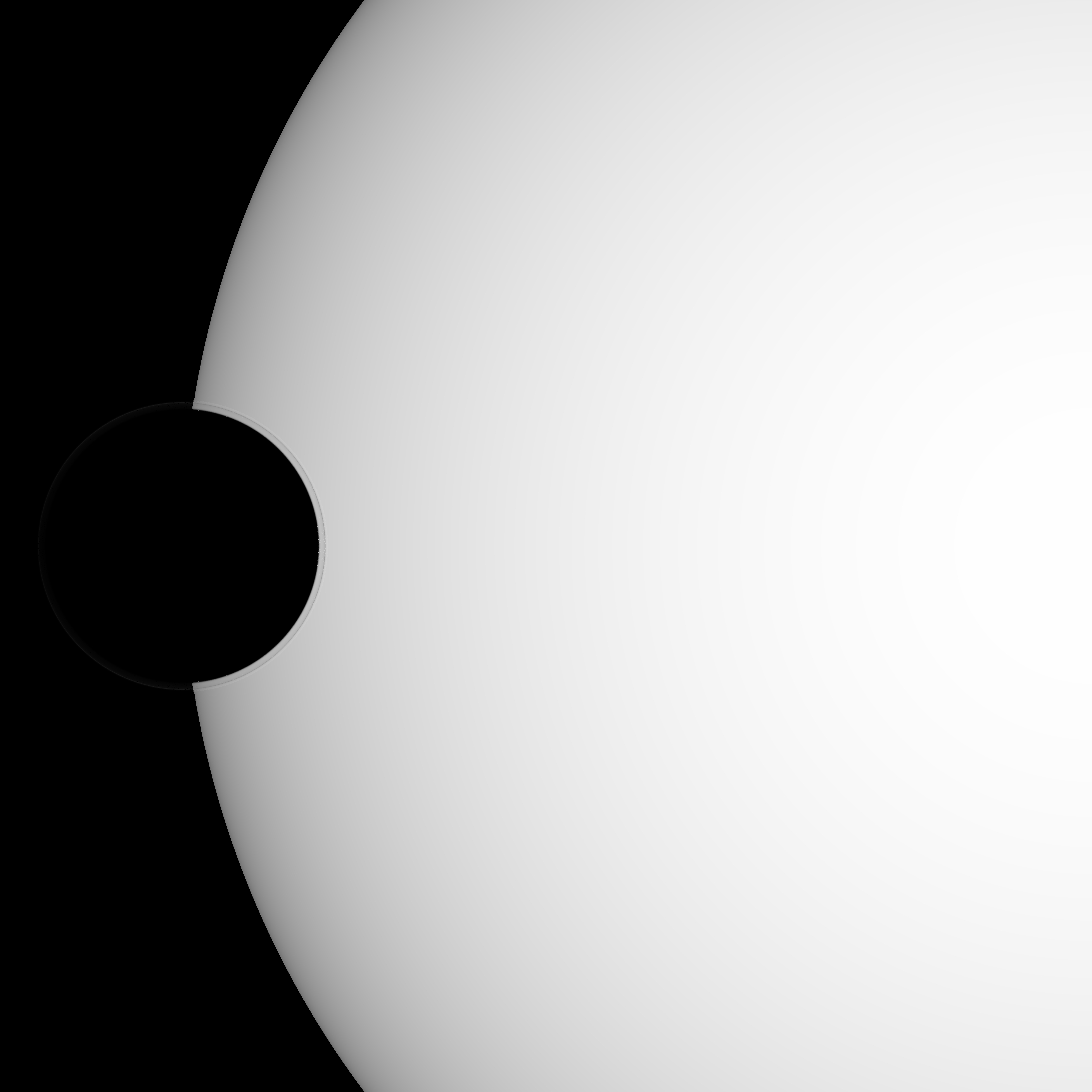} &
    \includegraphics[trim = 0mm 0mm 0mm 0mm, clip, width=3.1in]{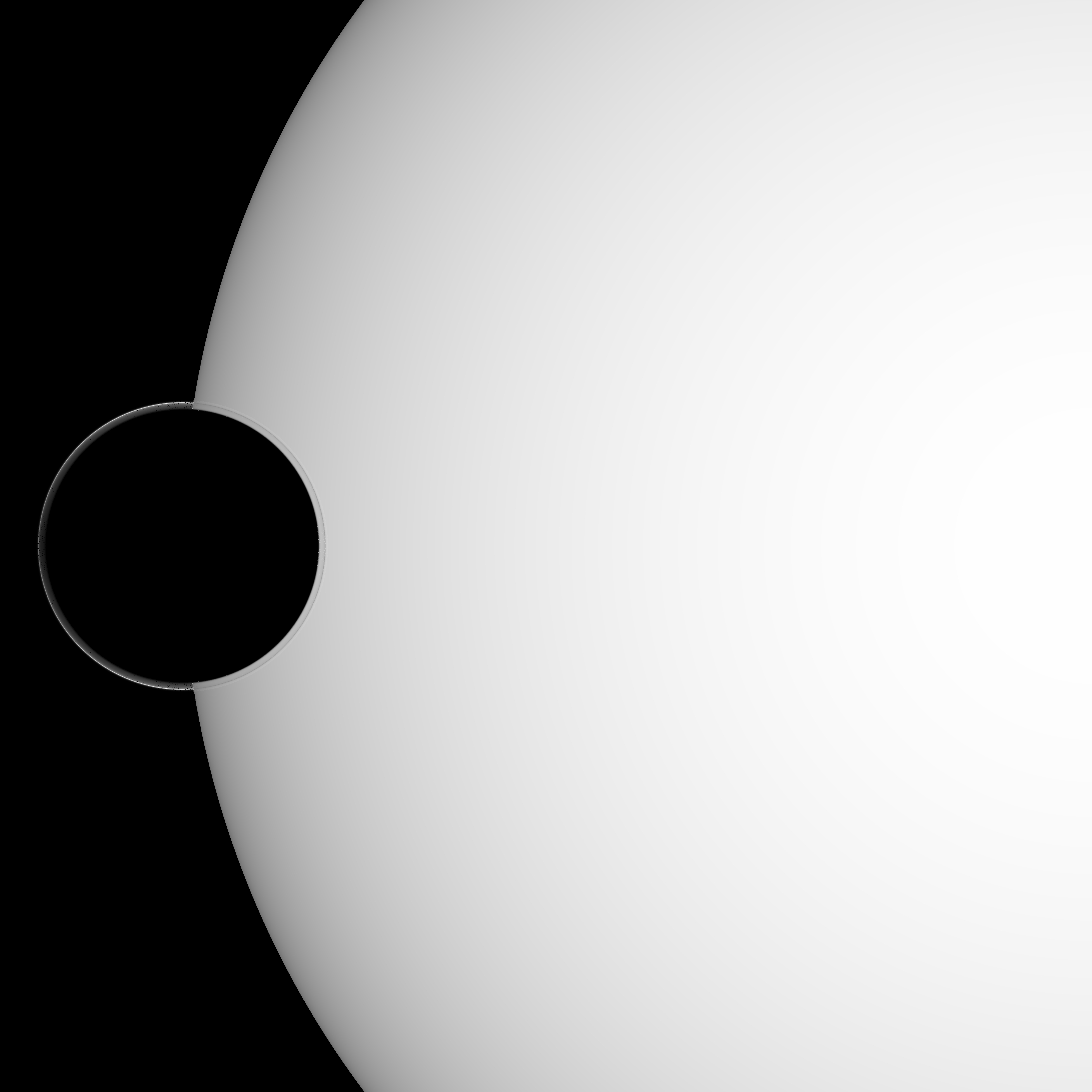} \\
  \end{tabular}
  \caption{Result of a full Monte Carlo transit spectrum simulation prior to solid angle integration. 
                The wavelength is red-visible (0.8~$\mu$m), hot Jupiter-like conditions are assumed, 
                refraction and Rayleigh scattering are included, and a forward scattering 
                ($g=0.9$) haze is placed above the 1~$\mu$bar pressure level. The right  
                sub-figure has the surface brightness off the stellar disk enhanced by a factor of 10.}
  \label{fig:simulation}
\end{figure}
\clearpage
\begin{figure}
  \centering
  \begin{tabular}{cc}
    \includegraphics[trim = 0mm 0mm 0mm 0mm, clip, width=3.1in]{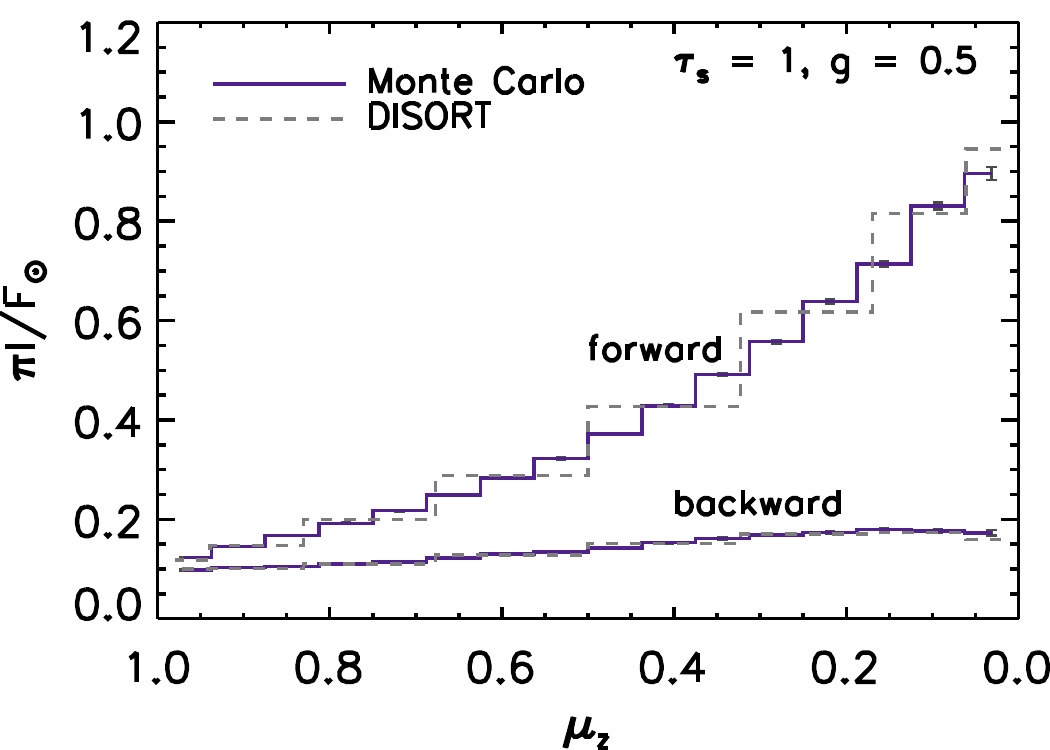} &
    \includegraphics[trim = 0mm 0mm 0mm 0mm, clip, width=3.1in]{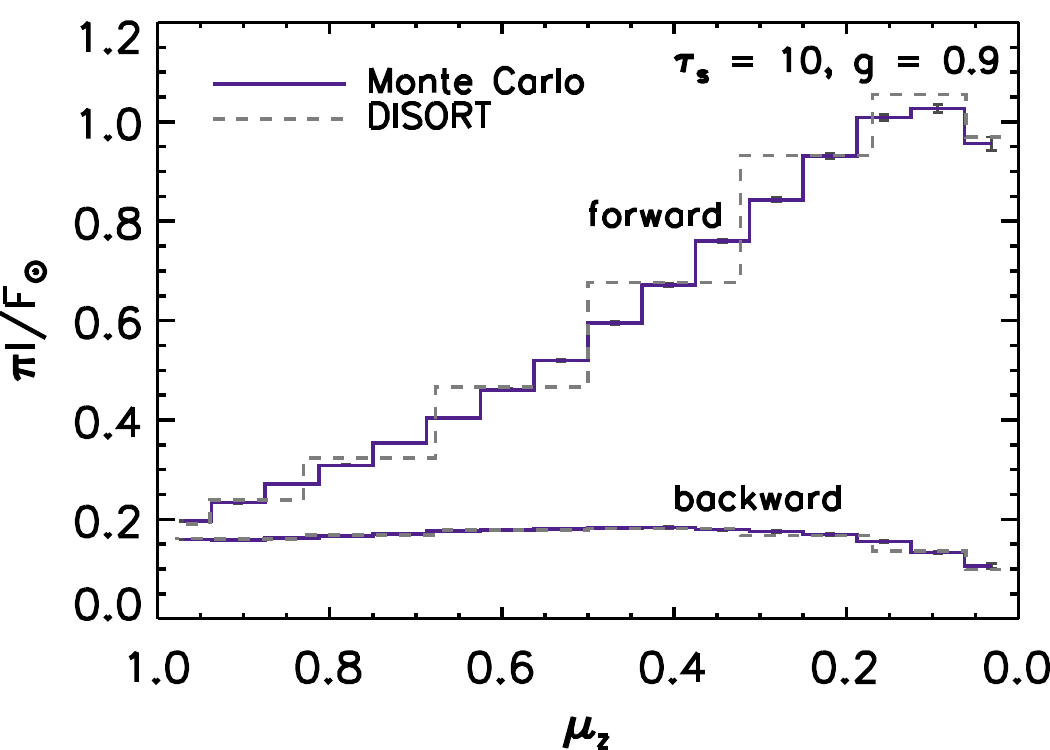} \\
  \end{tabular}
  \caption{Comparison of the output radiances (scaled by the incident flux) from our Monte Carlo 
                routine, in the plane-parallel limit, against those from the {\tt DISORT} radiative transfer 
                model \citep{stamnesetal1988}.  Clouds with a given scattering optical depth and 
                asymmetry parameter are placed over a black surface, and a Henyey-Greenstein phase 
                function is assumed.  Results are shown as a function of the observer zenith angle for 
                both the forward and backward scattering azimuths in the plane of the incident light 
                source.}
  \label{fig:mc_valid}
\end{figure}
\clearpage
\begin{figure}
  \centering
  \includegraphics[trim = 0mm 0mm 0mm 0mm, clip, width=6.2in]{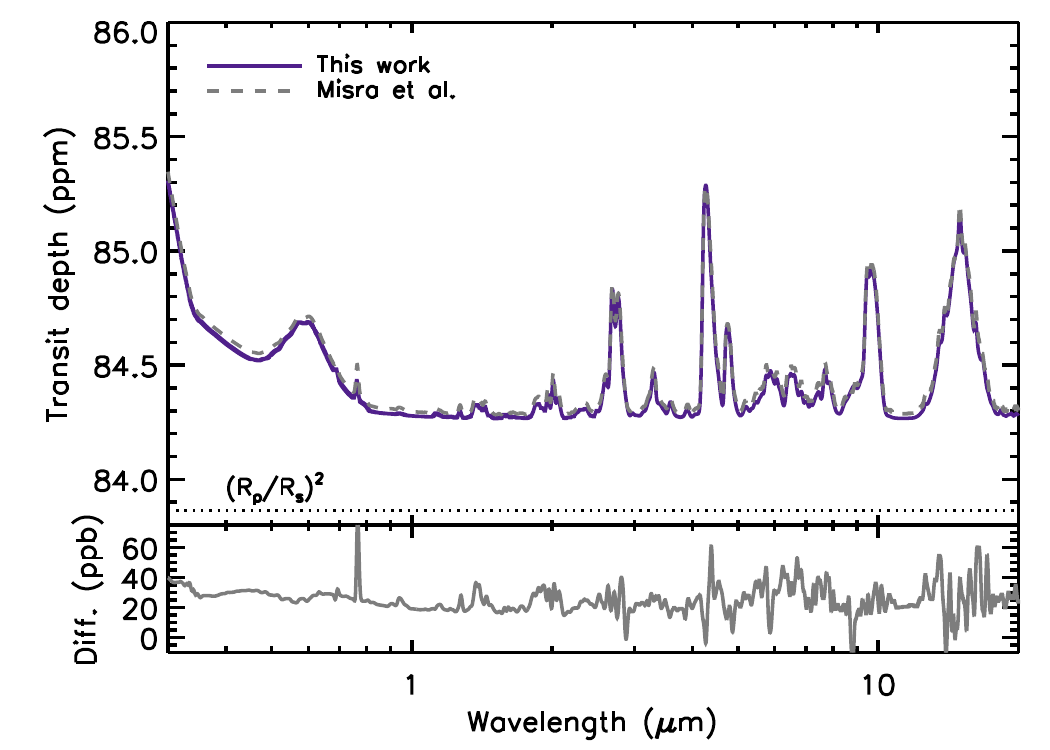}
  \caption{Comparison between our transit spectrum model and the refracting model of 
                \citet{misraetal2014} for a standard Earth atmospheric model 
                \citep{mcclatcheyetal1972}.  { Differences between the models are 
                shown in the lower sub-panel, and are typically within 60 ppb.}}
  \label{fig:ref_valid}
\end{figure}
\clearpage
\begin{figure}
  \centering
  \includegraphics[trim = 0mm 0mm 0mm 0mm, clip, width=6.2in]{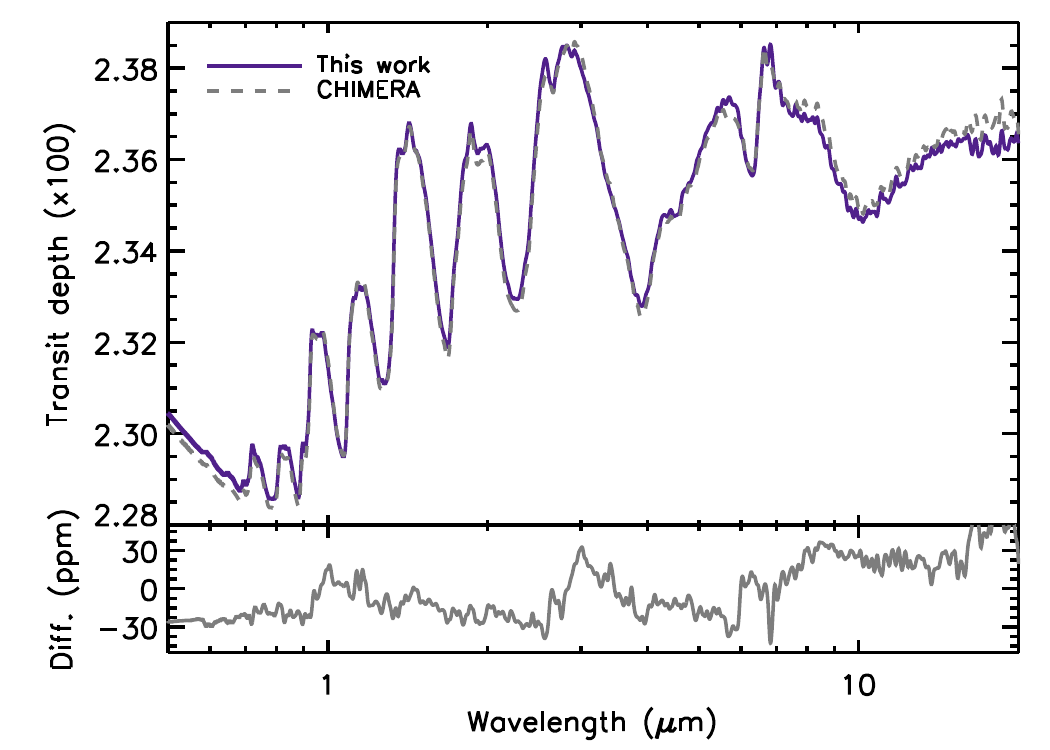}
  \caption{Comparison between our transit spectrum model, in the geometric limit, and 
                output from the CHIMERA retrieval suite \citep{lineetal2013a} for a standard hot 
                Jupiter-like model atmosphere described in the text.  { Differences between the models are 
                shown in the lower sub-panel, and are typically within 30 ppm.}}
  \label{fig:geom_valid}
\end{figure}
\clearpage
\begin{figure}
  \centering
  \includegraphics[trim = 0mm 0mm 0mm 0mm, clip, width=6.2in]{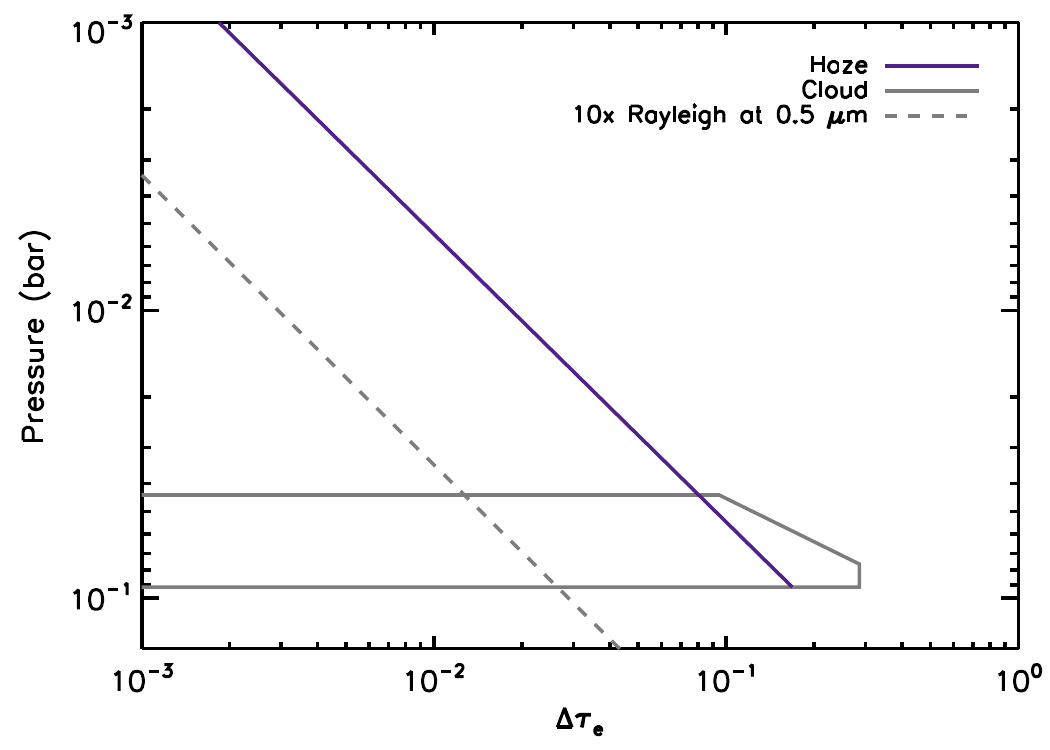}
  \caption{Layer differential extinction optical depths for a nominal haze-like model and a 
                condensate cloud-like model \citep[after][]{ackerman&marley2001} used when 
                computing example path distributions for a hot Jupiter-like atmosphere.  { For comparison, 
                the layer differential Rayleigh scattering optical depths (at 0.5~$\mu$m) for out hot 
                Jupiter-like atmosphere are also shown.}}
  \label{fig:demo_dtau}
\end{figure}
\clearpage
\begin{figure}
  \centering
  \includegraphics[trim = 0mm 0mm 0mm 0mm, clip, width=6.2in]{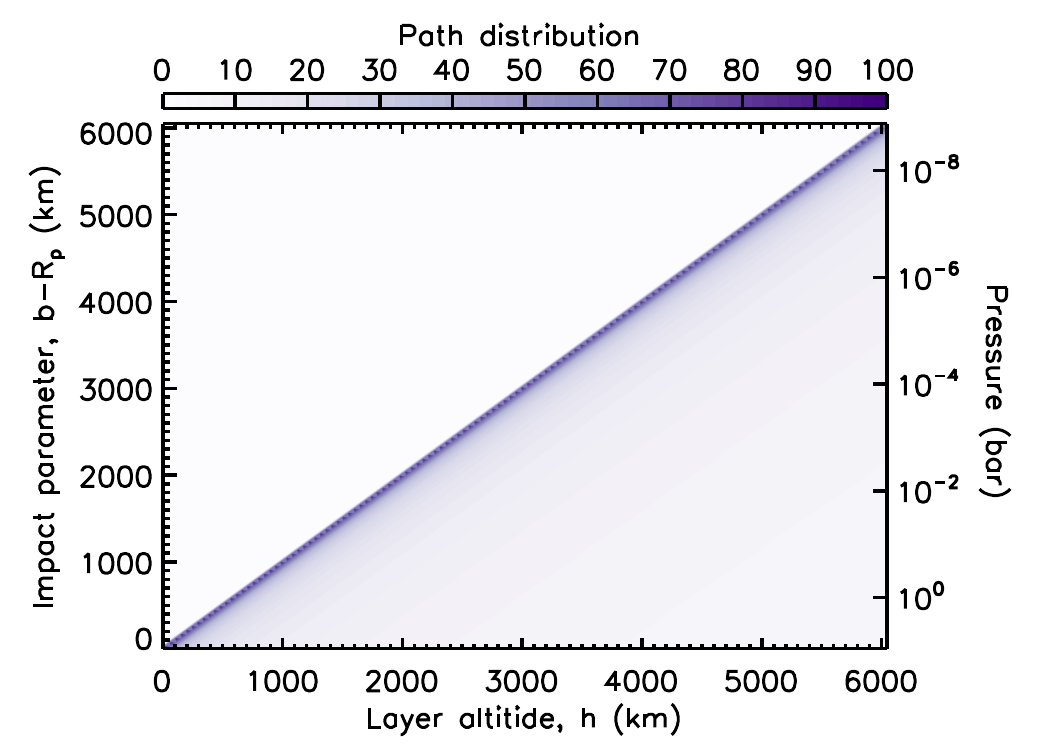}
  \caption{The path distribution, $\mathcal{P}_{b}(h)$, in the geometric limit for a hot 
                Jupiter-like atmosphere.  Altitudes are relative to the 10-bar planetary radius at 
                $1.16R_{\rm{J}}$.  Darker colors indicate larger path distribution values, which implies a 
                larger enhancement of the vertical differential optical depth.  The largest enhancements occur 
                when a ray is passing near the horizontal for a layer, which is where $b-R_{\rm{p}}$ is roughly 
                equal $h$.}
  \label{fig:demo_geom}
\end{figure}
\clearpage
\begin{figure}
  \centering
  \begin{tabular}{cc}
    \includegraphics[trim = 0mm 0mm 0mm 0mm, clip, width=3.1in]{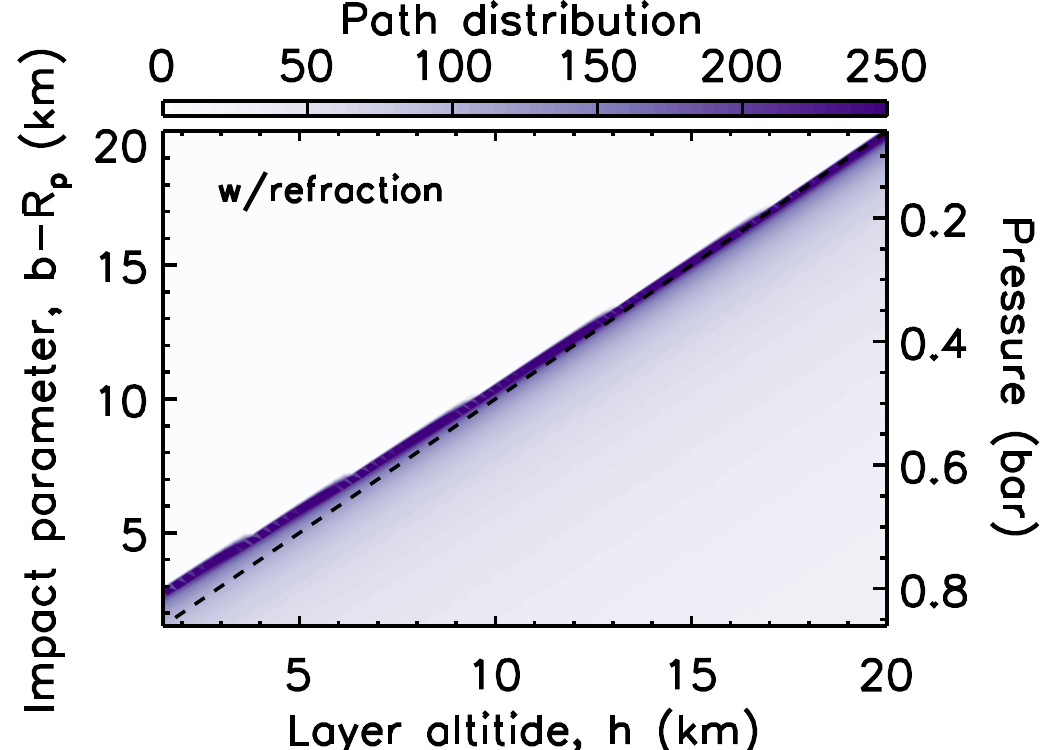} &
    \includegraphics[trim = 0mm 0mm 0mm 0mm, clip, width=3.1in]{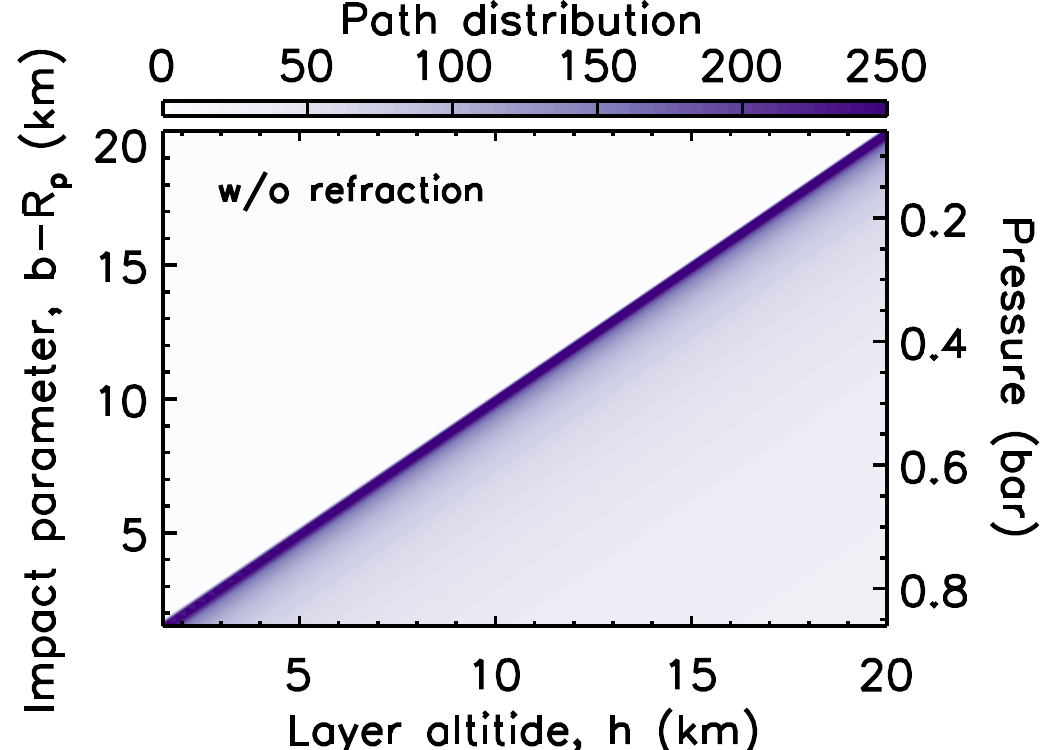} \\
  \end{tabular}
  \caption{The path distribution, $\mathcal{P}_{b}(h)$, for a cloud-free Earth atmosphere 
                that includes the effects of refraction.  A dashed line is given along the diagonal.  
                The right sub-figure shows, for comparison, the path distribution without 
                refraction.  The path distribution for rays with impact parameter altitudes below 
                1.6~km are not shown as these rays strike the surface.  Rays with impact parameter 
                altitudes smaller than about 10~km are deflected downward due to refraction, and 
                thus probe deeper atmospheric layers than in the geometric case.  Here the peak of 
                the path distribution is also shifted to slightly lower altitudes.}
  \label{fig:demo_refract}
\end{figure}
\clearpage
\begin{figure}
  \centering
  \includegraphics[trim = 0mm 0mm 0mm 0mm, clip, width=6.2in]{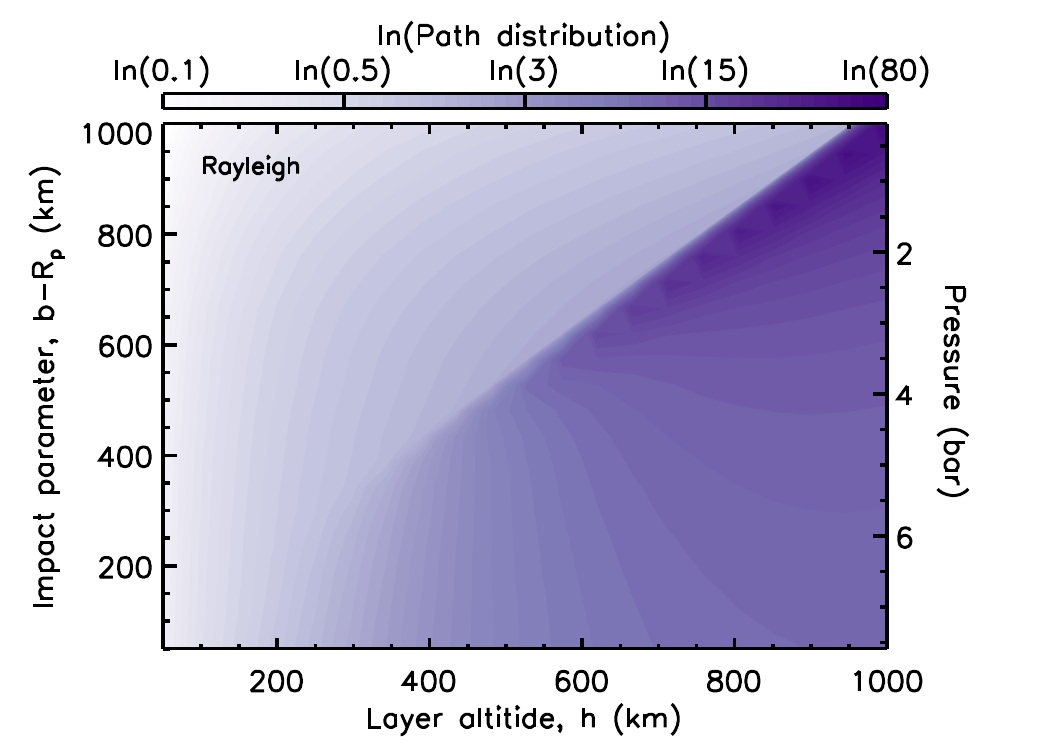}
  \caption{The average path distribution, $\bar{\boldsymbol{\mathcal{P}}}$, including 
                molecular Rayleigh scattering at 0.55~$\mu$m wavelength for a hot Jupiter-like 
                atmosphere.  Photons (or rays) with impact parameter altitudes above about 1000~km 
                only probe low pressure regions of the atmosphere, and experience relatively little 
                Rayleigh scattering.  Photons with impact parameters below about 400--600~km 
                experience substantial Rayleigh scattering, and are typically scattered before reaching 
                deeper atmospheric layers.  Thus, rays/photons with low impact parameter altitudes 
                have small path distributions at depth and relatively large path distributions aloft.}
  \label{fig:demo_ray}
\end{figure}
\clearpage
\begin{figure}
  \centering
  \begin{tabular}{cc}
    \includegraphics[trim = 0mm 0mm 0mm 0mm, clip, width=3.1in]{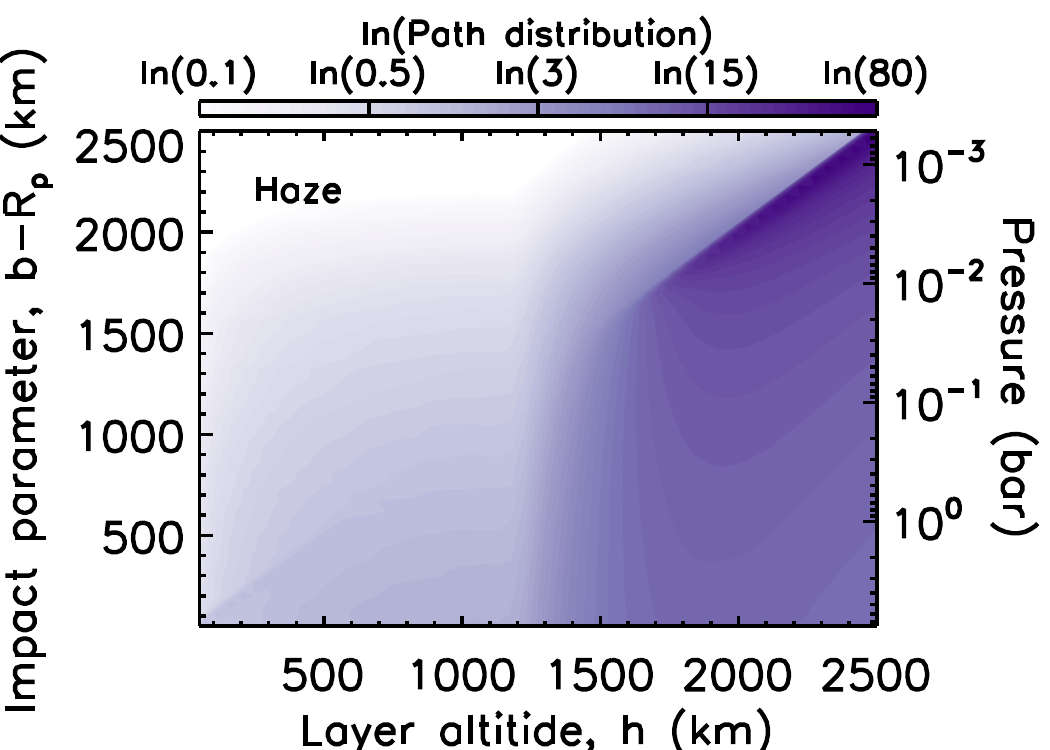} &
    \includegraphics[trim = 0mm 0mm 0mm 0mm, clip, width=3.1in]{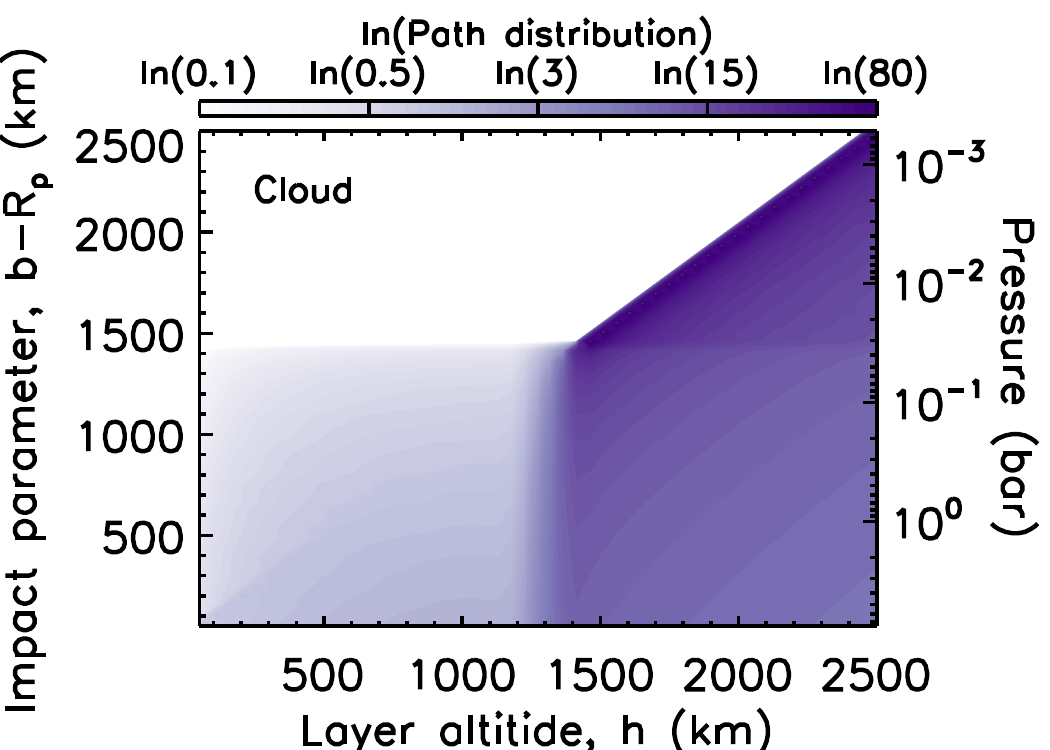} \\
  \end{tabular}
  \caption{The average path distribution, $\bar{\boldsymbol{\mathcal{P}}}$, for a hot Jupiter-like   
                 atmosphere where multiple scattering in an upper atmospheric haze (left) and  
                condensate cloud (right) are considered.  Aerosol optical depth distributions are  
                shown in Figure~\ref{fig:demo_dtau}, the single-scattering albedo is unity, and the 
                asymmetry parameter is taken as forward scattering ($g=0.9$).  The haze case 
                resembles the Rayleigh scattering case in Figure~\ref{fig:demo_ray} due to the 
                power-law distribution of haze scattering optical depth.  The condensate cloud 
                case is identical to the geometric case for rays/photons with impact parameters 
                above the cloud.  For impact parameters below (or in) the cloud, scattering prevents 
                photons from reaching deeper atmospheric layers, so that the path distribution is 
                distrinctly different above versus below the cloud.}
  \label{fig:demo_clouds}
\end{figure}
\clearpage
\begin{figure}
  \centering
  \begin{tabular}{c}
    \includegraphics[trim = 0mm 0mm 0mm 0mm, clip, width=3.1in]{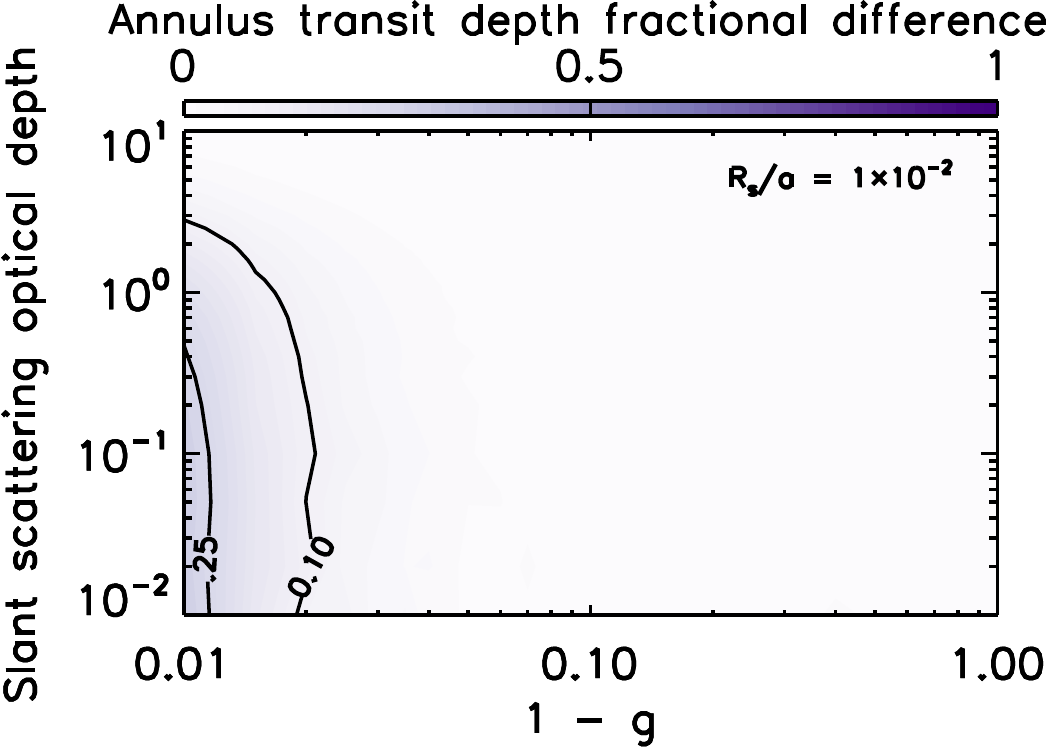} \\
    \includegraphics[trim = 0mm 0mm 0mm 0mm, clip, width=3.1in]{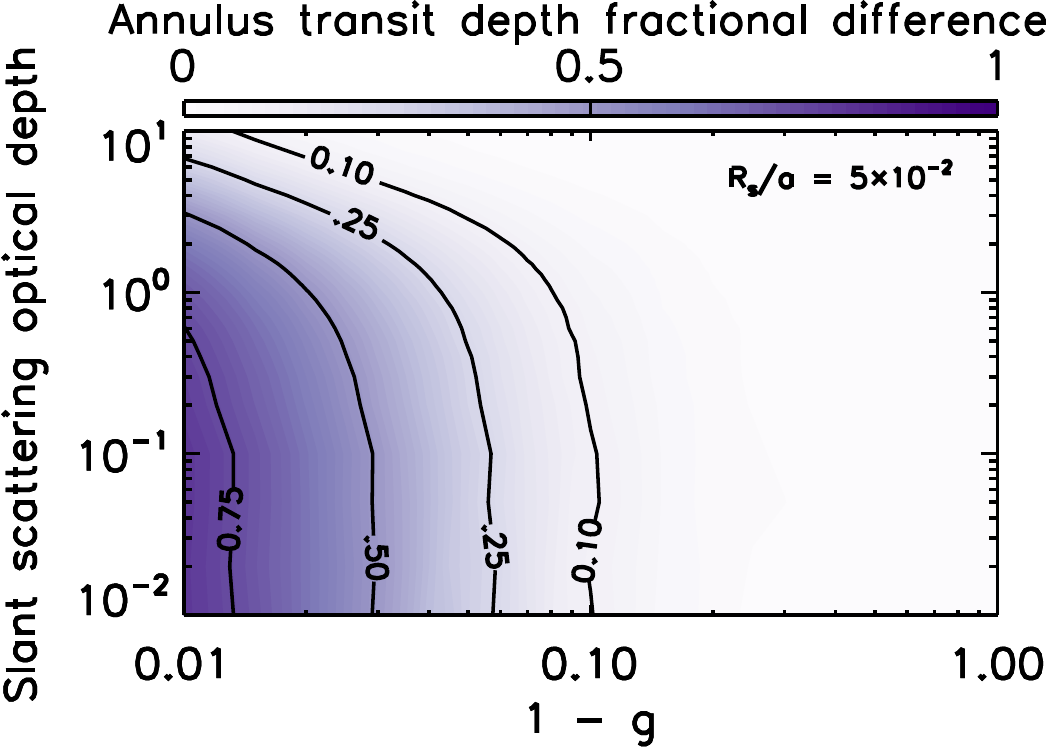} \\
    \includegraphics[trim = 0mm 0mm 0mm 0mm, clip, width=3.1in]{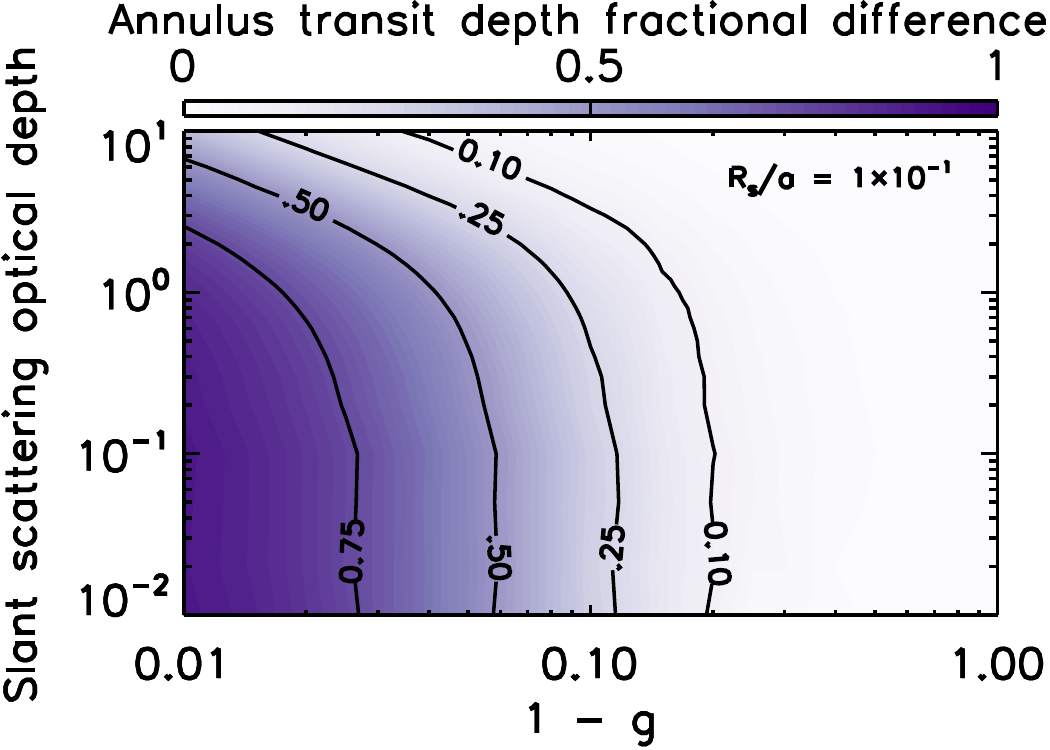} \\
  \end{tabular}
  \caption{Relative difference between the transit depth due to a single annulus in the 
                geometric limit versus a model that includes multiple scatterings.  Sub-figures are 
                for different angular sizes of the host star as seen from the planet, and results are 
                given as a function of scattering slant optical depth within the annulus and the 
                scattering asymmetry parameter.}
  \label{fig:tdepth_var}
\end{figure}
\clearpage
\begin{figure}
  \centering
  \begin{tabular}{cc}
    \includegraphics[trim = 0mm 0mm 0mm 0mm, clip, width=3.1in]{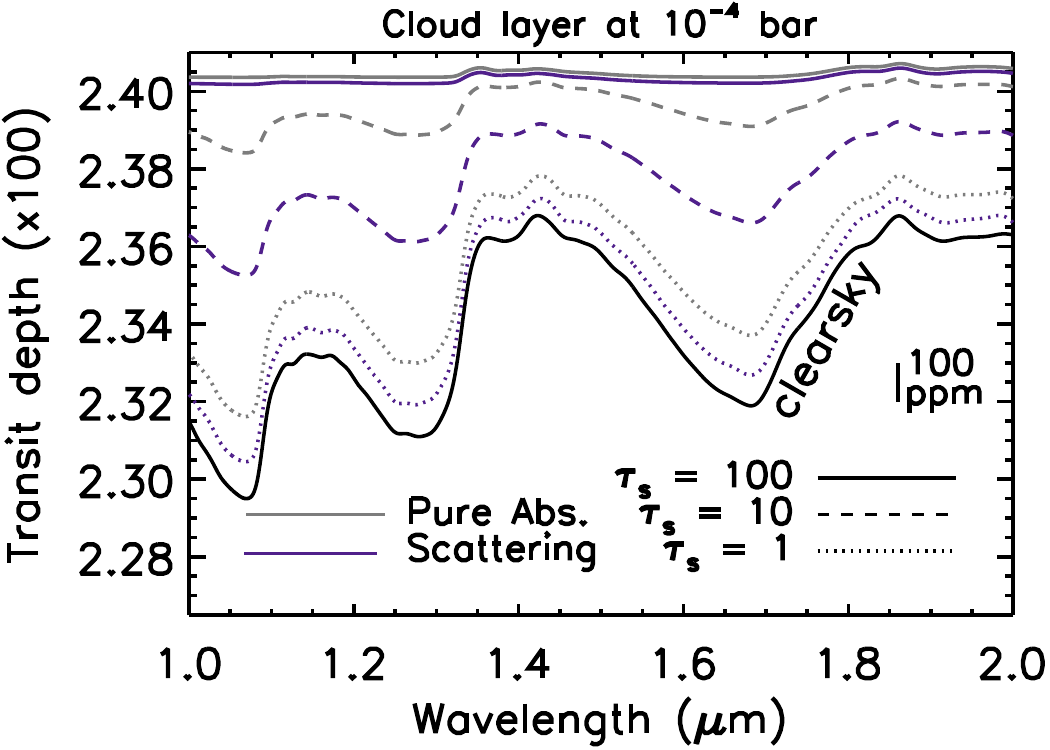} &
    \includegraphics[trim = 0mm 0mm 0mm 0mm, clip, width=3.1in]{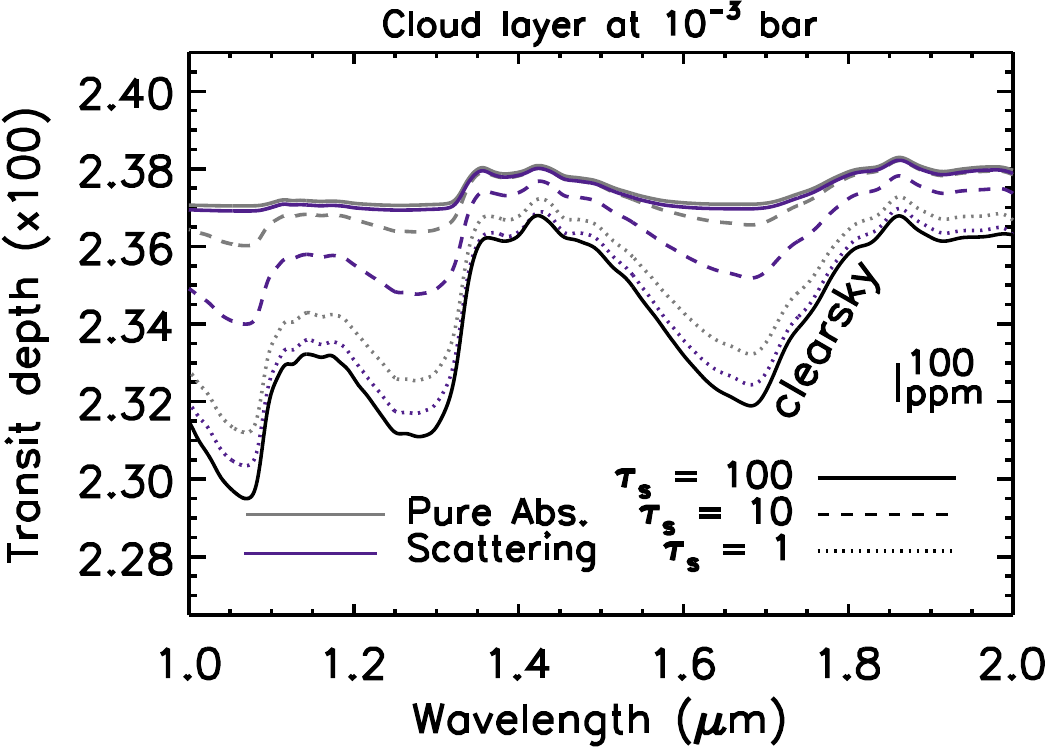} \\
    \includegraphics[trim = 0mm 0mm 0mm 0mm, clip, width=3.1in]{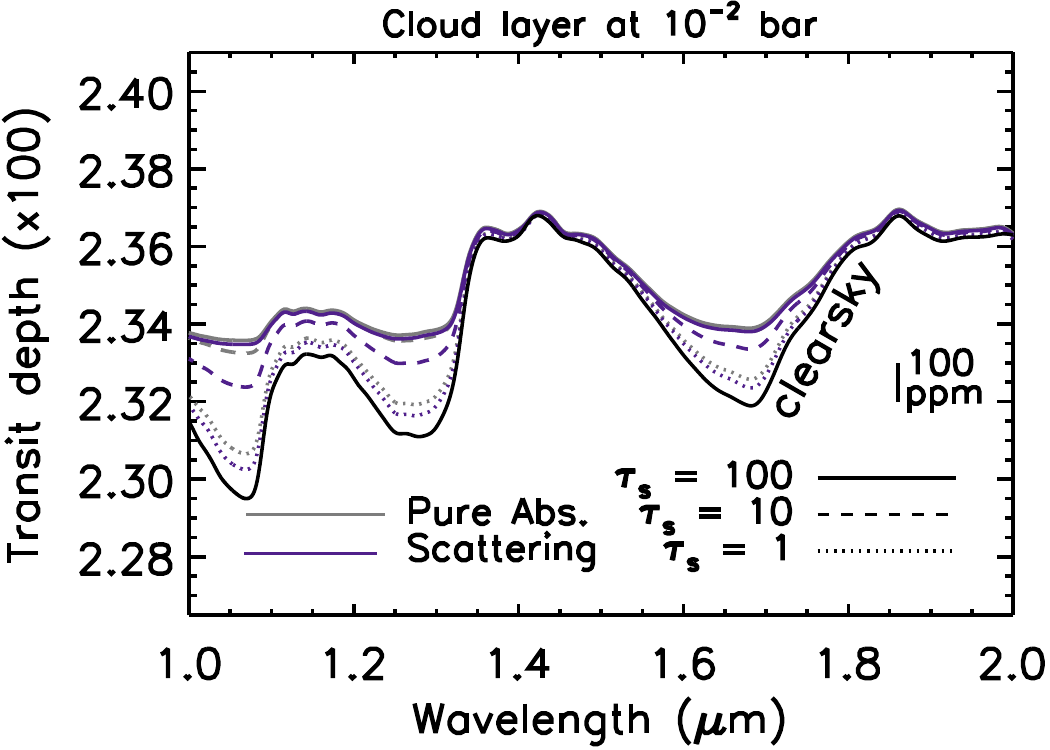} &
    \includegraphics[trim = 0mm 0mm 0mm 0mm, clip, width=3.1in]{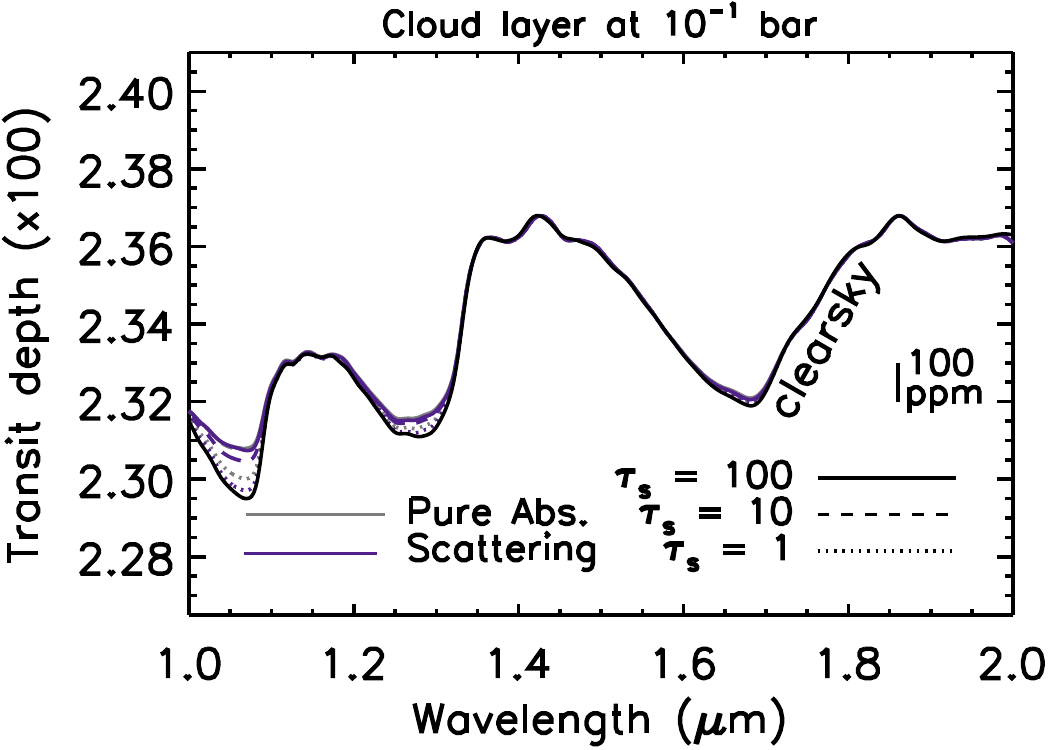} \\
  \end{tabular}
  \caption{Near-infrared transit spectra for a hot Jupiter-like planet.  Different panels are for cases where 
                a vertically thin cloud is placed at the indicated pressure level.  Different line styles indicate cases 
                with different slant scattering optical depths for the cloud.  Gray lines assume that all optical depth 
                is absorption optical depth, while purple lines include realistic scattering.  The clearsky limit is shown 
                in black.  Aerosol optical properties are assumed to be gray, a Henyey-Greenstein scattering phase 
                function is used, and the aerosols are taken to be forward scattering, with $g=0.95$.}
  \label{fig:tdepth_thincloud_g0.95}
\end{figure}
\clearpage
\begin{figure}
  \centering
  \begin{tabular}{cc}
    \includegraphics[trim = 0mm 0mm 0mm 0mm, clip, width=3.1in]{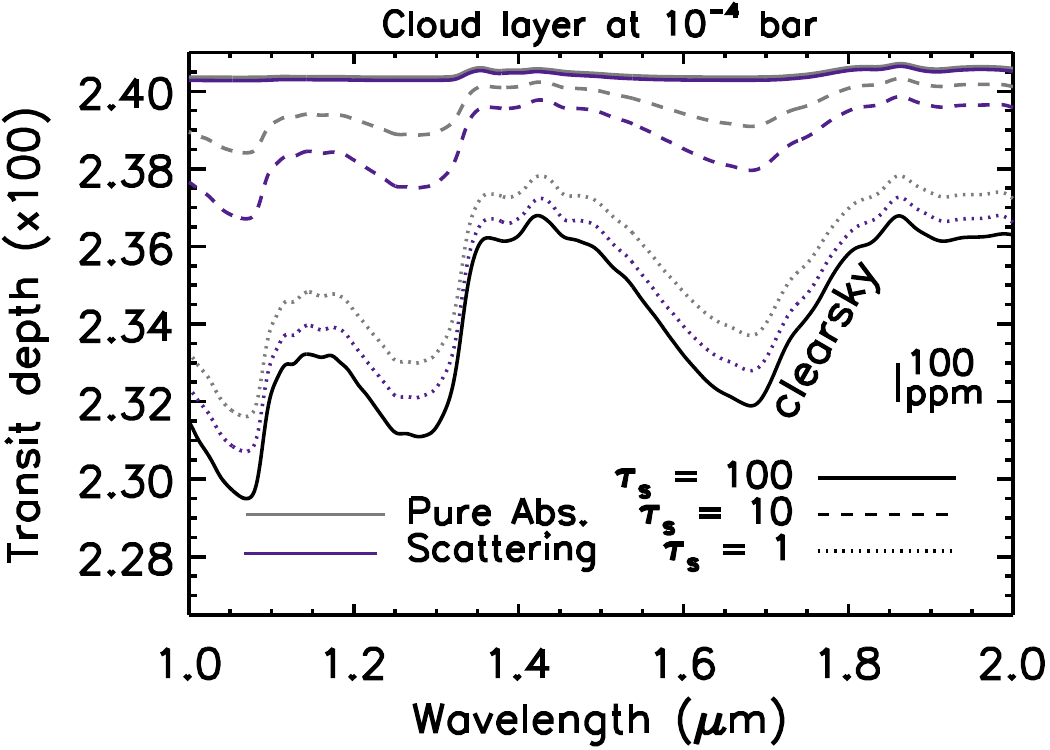} &
    \includegraphics[trim = 0mm 0mm 0mm 0mm, clip, width=3.1in]{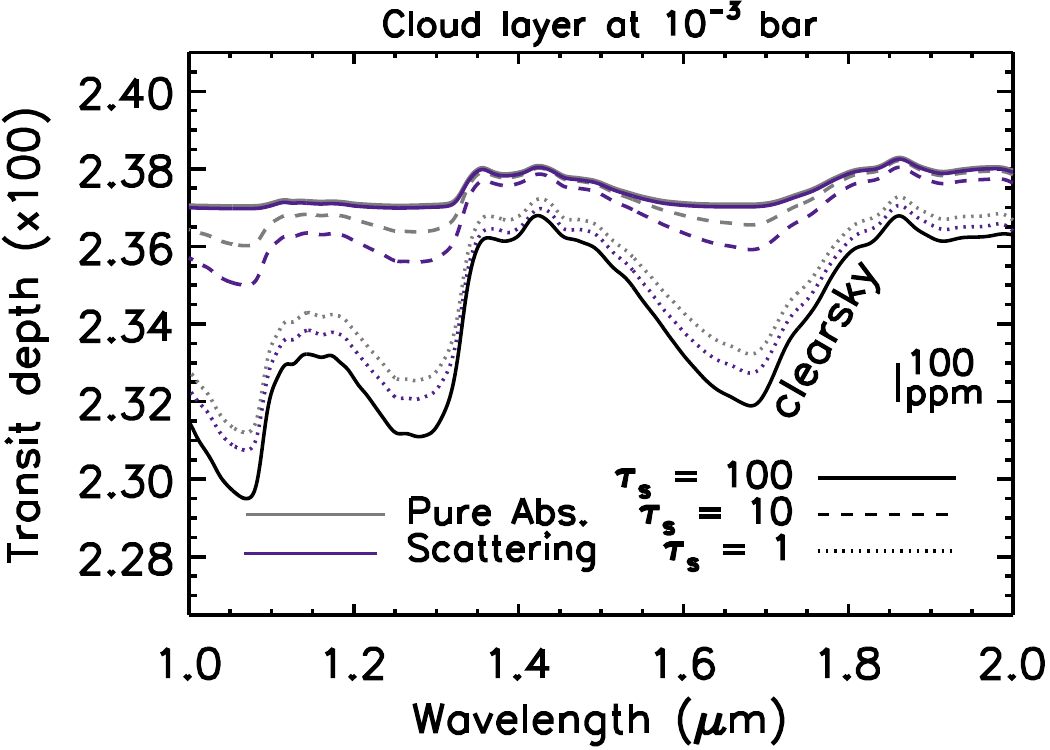} \\
    \includegraphics[trim = 0mm 0mm 0mm 0mm, clip, width=3.1in]{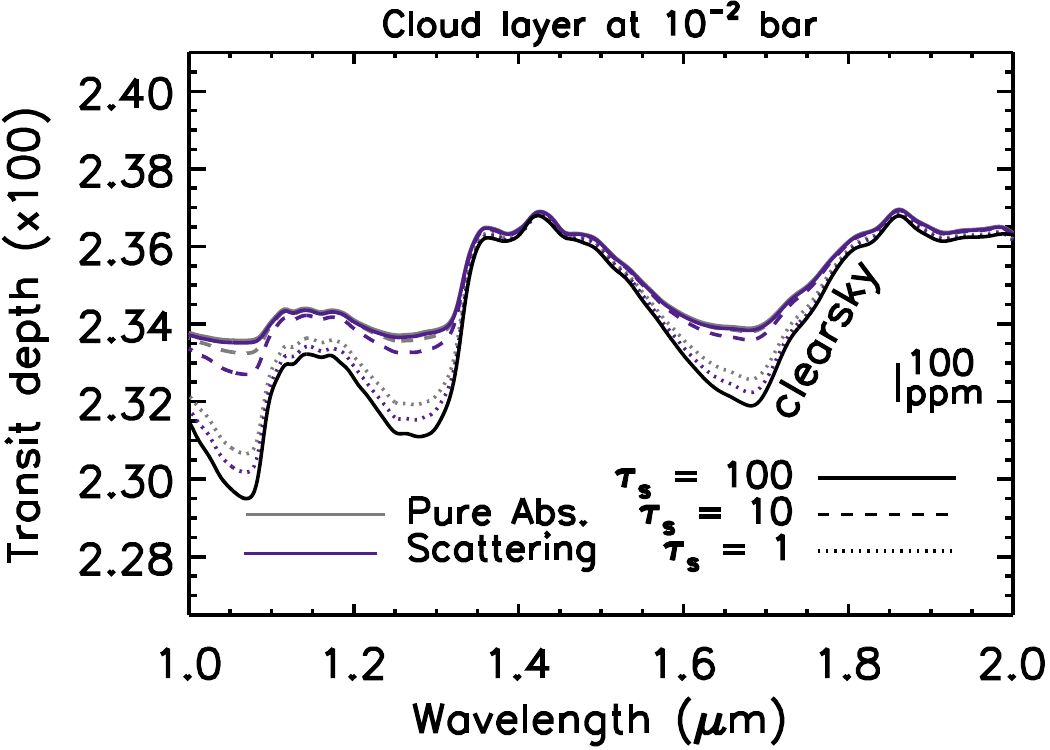} &
    \includegraphics[trim = 0mm 0mm 0mm 0mm, clip, width=3.1in]{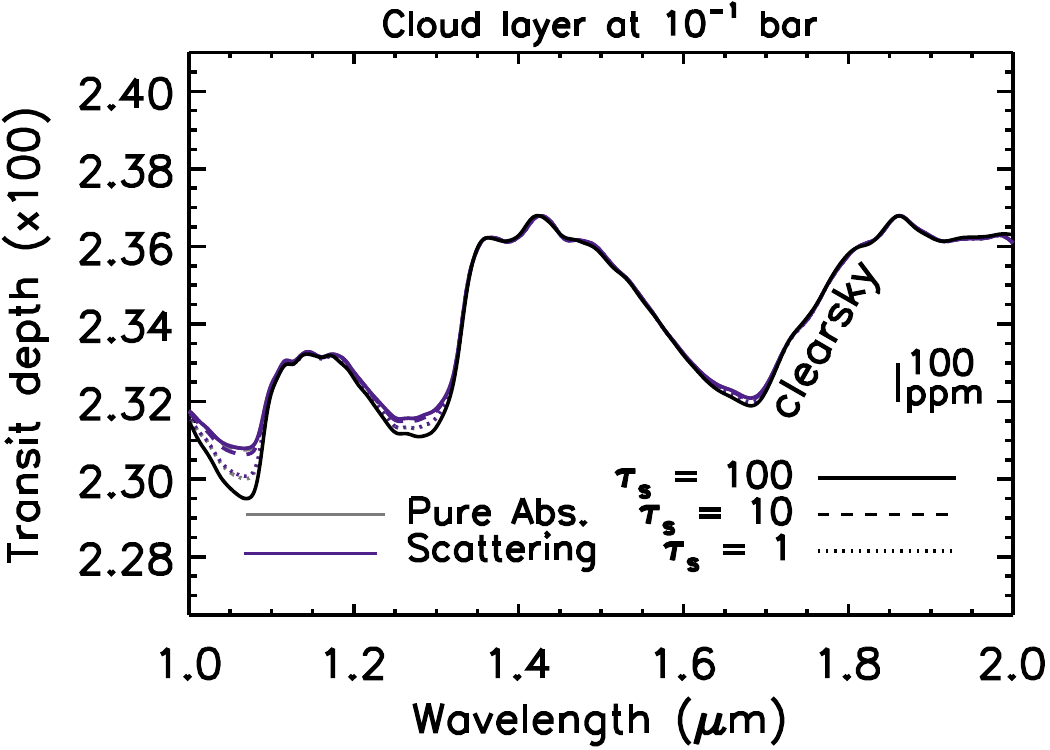} \\
  \end{tabular}
  \caption{The same as Figure~\ref{fig:tdepth_thincloud_g0.95} except with $g=0.90$.}
  \label{fig:tdepth_thincloud_g0.90}
\end{figure}
%
\end{document}